\documentclass[a4paper,fleqn]{cas-dc}

\usepackage[numbers]{natbib}
\usepackage{algorithm}
\usepackage{algpseudocode}
\makeatletter
\renewcommand{\theHALG@line}{\thealgorithm.\arabic{ALG@line}}
\makeatother

\usepackage{amsmath}
\usepackage{threeparttable}
\usepackage{graphicx}
\usepackage{booktabs}
\usepackage{multirow}

\usepackage{hyperref}
\usepackage{array}
\usepackage{booktabs}
\usepackage{caption}
\usepackage{graphicx}
\usepackage{tikz}

\usepackage{xurl}

\newcommand{\customcirc}[1]{%
\ifthenelse{\equal{#1}{emptycirc}}{%
\tikz\draw[black, line width=0.45pt] (0,0) circle (0.65ex);%
}{%
\ifthenelse{\equal{#1}{halfcirc}}{%
\tikz\fill[black] (0,0) circle (0.65ex);%
}{%
\ifthenelse{\equal{#1}{fullcirc}}{%
\tikz\fill[gray!55] (0,0) circle (0.65ex);%
}{}%
}%
}%
}
\makeatletter
\AddToHook{env/algorithmic/before}{\def\@currentcounter{ALG@line}}
\makeatother
\usepackage[capitalize]{cleveref} % 'capitalize' makes it "Line" instead of "line"
\crefalias{ALG@line}{line}

\def\tsc#1{\csdef{#1}{\textsc{\lowercase{#1}}\xspace}}
\tsc{WGM}
\tsc{QE}
\begin{document}
\let\WriteBookmarks\relax
\def\floatpagefraction{1}
\def\textfraction{.001}

% Short title
\shorttitle{LCC-LLM}    

% Short author

% Main title of the paper
\title [mode = title]{LCC-LLM: Leveraging Code-Centric Large Language Models for Malware Attribution}  

% Title footnote mark
% eg: \tnotemark[1]

% First author
%
% Options: Use if required
% eg: \author[1,3]{Author Name}[type=editor,
%       style=chinese,
%       auid=000,
%       bioid=1,
%       prefix=Sir,
%       orcid=0000-0000-0000-0000,
%       facebook=<facebook id>,
%       twitter=<twitter id>,
%       linkedin=<linkedin id>,
%       gplus=<gplus id>]

% Address/affiliation
\author[1]{Christopher G. Pedraza Pohlenz}

\author[1]{Hassan Jalil Hadi}[orcid=0000-0001-7746-344X]
\cormark[1]
\ead{hassan.hadi@kaust.edu.sa}

\author[1]{Ali Hassan}

\author[1]{Ali Shoker}

\affiliation[1]{
    organization={CyberSaR, King Abdullah University of Science and Technology},
    city={Thuwal},
    postcode={23955-6900},
    state={Makkah Province},
    country={Saudi Arabia}
}

\cortext[1]{Corresponding author}

% For a title note without a number/mark
%\nonumnote{}

% Here goes the abstract
\begin{abstract}
Large Language Models (LLMs) are increasingly being explored for malware analysis; however, current LLM-based malware attribution remains limited by unsupported indicators and insufficient code-level grounding for identifying malicious and vulnerable code segments. To address these limitations, this research introduces LCC-LLM, a code-centric benchmark dataset and evidence-grounded intelligent framework for malware attribution and multi-task static malware analysis. The proposed LCCD \footnote{\fntext[1]{LCC-LLM refers to the proposed framework, and LCCD refers to the benchmark dataset.}} dataset contains approximately 34K PE samples processed through a large-scale reverse-engineering pipeline and represented using decompiled C code, assembly code, CFG/FCG artifacts, hexadecimal data, PE metadata, suspicious API evidence, and structural features. Beyond dataset construction, LCC-LLM integrates LangGraph-orchestrated static analysis with multi-source cybersecurity knowledge. This integration supports evidence-grounded malware reasoning. The framework also employs a seven-layer retrieval-augmented generation pipeline. It uses Chain-of-Verification for IoC validation. In addition, a multi-dimensional quality gate is applied to improve factual reliability and analyst-oriented decision support. Curriculum-ordered instruction data is used to fine-tune DeepSeek-R1-Distill-Qwen-14B and Qwen3-Coder-30B-A3B using QLoRA. Evaluation across 43 malware-analysis task types achieves an average semantic similarity of 0.634, with the highest task-level performance observed in structured report generation, IoC extraction, vulnerability assessment, malware configuration extraction, and malware class detection. In a real-world case-study evaluation using MalwareBazaar samples, the grounded pipeline achieves a 10/10 structured analysis pass rate, producing CFG/FCG evidence, MITRE ATT\&CK mappings, detection guidance, and analyst-ready reports. These results show that code-centric representations, retrieval grounding, and verification-guided reasoning improve the reliability and operational usefulness of LLM-assisted malware attribution.
\end{abstract}

% Research highlights
\begin{highlights}
\item We introduce LCCD, a code-centric benchmark dataset for Windows PE malware attribution.

\item LCC-LLM combines LangGraph-orchestrated static analysis, seven-layer RAG, CoT, CoVe, and quality-gated reasoning.

\item The framework fine-tunes code-oriented LLMs for attribution, malicious code segmentation, vulnerability analysis, and reporting.

\item A chatbot-based expert-system prototype supports interactive malware triage and analyst workflows.

\end{highlights}

%\nocite{*}

% Keywords
% Each keyword is seperated by \sep

\maketitle

% Main text
%~~~~~~~~~~~~~~~~~~~~~~~~~~~~~~~~~~~~~~~~~~~~~~~~~~~~~~~~~~~~~~~~~~~~~~~~~~~~~~~~~~~~~~~~
\section{Introduction}\label{sec:introduction}
Malware continues to pose a major threat to modern digital infrastructures, affecting enterprise networks, cloud services, critical systems, and Security Operations Centers (SOCs) \cite{ling2023adversarial, xenos2025cross}. As malware families evolve through packing, obfuscation, code reuse, and rapid variant generation, timely malware attribution has become an essential task for cyber threat intelligence \cite{geng2024survey}, incident response, and defensive decision-making \cite{hadi2025fcg}. Malware attribution is not limited to assigning a family label to a suspicious binary; it requires understanding malicious intent, identifying suspicious code segments, interpreting API behavior, mapping observed behaviors to known attack techniques, extracting Indicators of Compromise (IoCs) \cite{williamson2024malware}, and generating actionable guidance for analysts \cite{jelodar2025large}. Therefore, effective malware analysis systems must go beyond simple detection and provide evidence-grounded explanations that connect low-level reverse-engineering artifacts with higher-level threat intelligence.

Furthermore, traditional malware analysis methods commonly rely on static or dynamic features such as byte sequences, opcode patterns, imported APIs, PE metadata, control-flow features, sandbox traces, or manually engineered indicators \cite{or2019dynamic, ucci2019survey}. These approaches have been widely used for malware detection and family classification, but they often operate on isolated feature spaces and provide limited support for analyst-oriented reasoning. For example, a classifier may predict that a sample belongs to a particular malware family \cite{hadi2025fcg, egele2008survey}, but it may not explain which code structures, suspicious APIs, function relationships, or threat-intelligence indicators support that attribution \cite{kalgutkar2019code}. Similarly, static feature-based systems may identify suspicious imports or entropy values but fail to connect these observations with malicious code segments \cite{guo2023empirical, fang2022jstrong, patsakis2024assessing}, vulnerable code patterns \cite{ding2024vulnerability}, MITRE ATT\&CK techniques, or remediation actions. This gap limits their practical value in operational environments where analysts require traceable, interpretable, and actionable evidence.

In addition, LLMs have recently demonstrated strong capabilities in code understanding, natural language reasoning, and security-oriented analysis \cite{jelodar2025large, al2024exploring, feng2025llm}. These capabilities create new opportunities for malware analysis, particularly because reverse-engineering artifacts such as decompiled C code, assembly instructions, function-call relationships, API traces, and structured reports can be represented as language-like inputs \cite{boke2025digital}. However, directly applying general-purpose LLMs to malware attribution introduces several challenges. Current LLM-based malware analysis can produce unsupported indicators, hallucinated threat claims, incomplete code-level evidence, and responses that are not aligned with SOC workflows \cite{bai2026automated, jelodar2026llm4codere}. In malware analysis, such errors are not merely linguistic mistakes; hallucinated IoCs, incorrect MITRE mappings, or unsupported family claims can mislead incident response decisions and reduce analyst trust \cite{jelodar2025sban}. Therefore, LLM-based malware analysis requires code-centric grounding, verified threat intelligence, deterministic reverse-engineering evidence, and quality-controlled response generation.

To address these challenges, this paper presents LCC-LLM, a code-centric benchmark dataset and evidence-grounded LLM framework for malware attribution and multi-task static malware analysis. The proposed Large-Scale Code-Centric Dataset (LCCD) contains approximately 34K PE samples processed through a large-scale reverse-engineering pipeline. Building on this dataset, LCC-LLM integrates LangGraph-orchestrated static analysis, multi-source cybersecurity knowledge, retrieval-augmented generation, Chain-of-Verification (CoVe), Chain-of-Thought (CoT), and a multi-dimensional quality gate to generate evidence-grounded malware analysis outputs. The resulting curriculum-ordered instruction data is used to fine-tune code-oriented LLMs using Quantized Low-Rank Adaptation (QLoRA), enabling analyst-facing tasks such as malware attribution, malicious and vulnerable code analysis, IoC extraction, MITRE ATT\&CK mapping, detection guidance, and structured malware reporting. This design connects reverse-engineering artifacts, verified threat intelligence, and LLM-based reasoning into a unified framework for reliable malware attribution and analyst-assisted security investigation.

The main contributions of this study are summarized as follows:
\begin{itemize}
    \item Existing malware datasets often provide binaries, labels, API traces, or limited metadata, but they rarely unify reverse-engineering artifacts with explicit support for malicious and vulnerable code segment analysis. To address this gap, we introduce LCCD, a code-centric benchmark dataset of approximately 34K Windows PE samples represented through decompiled C code, assembly code, CFG/FCG artifacts, hexadecimal data, PE metadata, suspicious API evidence, and structural features.
    \item Current LLM-based malware analysis is limited by the lack of structured training data that connects low-level malware artifacts with analyst-oriented tasks. To address this limitation, we construct an LLM-ready instruction-tuning corpus covering malware attribution, malicious code identification, vulnerable code detection, suspicious API analysis, static behavior analysis, threat identification, detection guidance, and technique explanation.
    \item General-purpose LLMs can generate unsupported malware claims because their outputs are often not grounded in reverse-engineering evidence or verified threat intelligence. To address this problem, we propose LCC-LLM, an evidence-grounded framework that combines LangGraph-orchestrated static analysis, deterministic reverse-engineering tools, multi-source cybersecurity knowledge, seven-layer RAG, CoVe, CoT, and a multi-dimensional quality gate.
    \item In recent studies, LLM malware-analysis systems are evaluated only as offline text-generation models, without demonstrating usability in real analyst workflows. To address this gap, we fine-tune code-oriented LLMs using QLoRA and curriculum-ordered instruction data, evaluate the system across 43 malware-analysis task types, and demonstrate a real-time chatbot prototype for malware triage, IoC extraction, MITRE ATT\&CK mapping, detection guidance, and structured malware reporting.
\end{itemize}

The remainder of this paper is organized as follows. \cref{sec:related_work} reviews related work on malware datasets, non-LLM-based malware detection, LLM-based malware analysis and cybersecurity applications, and tool-augmented malware reasoning. \cref{sec:methodology_dataset} presents the construction of LCCD and its code-centric representation pipeline. \cref{sec:proposed_methodology} describes the LCC-LLM framework, including static analysis orchestration, knowledge enrichment, retrieval grounding, verification, and quality control. \cref{sec:experimental_results} presents model fine-tuning and experimental setup, and reports the evaluation results and real-world testing. Finally, \cref{sec:conclusion} concludes the paper and outlines future research directions.

%~~~~~~~~~~~~~~~~~~~~~~~~~~~~~~~~~~~~~~~~~~~~~~~~~~~~~~~~~~~~~~~~~~~~~~~~~~~~~~~~~~~~~~~~
\section{Theoretical background}\label{sec:theoretical_background}

\subsection{Curriculum Learning Schedule}\label{subsec:curriculum_learning_schedule}
This training strategy was first proposed by \citet{bengio_curriculum_2009}, who hypothesized that machine learning models could benefit from mimicking the incremental learning progression of human academic years. In human education, foundational concepts taught early on serve as schemas to help students process increasingly complex topics later. By incrementally raising the difficulty of the learning material, data that would initially be impossible to understand becomes manageable.

Our curriculum learning strategy is grounded in Cognitive Load Theory (CLT) \citep{sweller_cognitive_1988}, which posits that learning is bottlenecked by working memory capacity. To prevent this cognitive overload, training must balance three factors:
\begin{enumerate}
    \item\textbf{Intrinsic load:} The inherent complexity of the material or task. For our model, this corresponds
to the difficulty score of each training sample.
    \item\textbf{Extraneous load:} Distractions or noise. We minimize this through careful
data preprocessing, consistent formatting, and structured task templates.
    \item\textbf{Germane load:} Productive effort of schema formation. We maximize it by ensuring each stage builds upon previously acquired knowledge.
\end{enumerate}

While multi-stage curriculum learning---where distinct difficulty tiers are trained sequentially with independent learning rate schedules and checkpointing---is theoretically appealing, it introduces significant engineering constraints when combined with DeepSpeed ZeRO-2. As discussed in \cref{subsec:deepspeed}, ZeRO-2 works by partitioning optimizer states and gradients across GPUs. However, having to reinitialize the DeepSpeed engine for each new curriculum stage leads to the loss of optimizer states. Additionally, it could cause gradient synchronization failures or memory fragmentation.

To preserve training stability while maintaining the pedagogical benefits of curriculum learning, we implement a \textbf{continuous difficulty-sorted curriculum}. Let the SFT dataset be defined as an ordered sequence $S = (s_1, s_2, \dots, s_N)$ with $N=113{,}418$, where each instance $s_i = (x_i, y_i, d_i)$ consists of the input $x_i$, the target $y_i$, and the difficulty score $d_i$. Formally, $d_i$ is computed as a weighted linear combination of distinct domain-specific complexity metrics extracted from the sample. Given a set of $K$ feature scoring functions $\phi_k(x_i)$ and their corresponding weights $\omega_k$, the difficulty is defined as:
\begin{equation}
d_i = \sum_{k=1}^{K} \omega_k \phi_k(x_i)
\end{equation}

subject to the constraint that the weights sum to unity ($\sum_{k=1}^{K} \omega_k = 1$). To enforce the curriculum, $S$ is strictly sorted such that $d_i \leq d_{i+1}$ for all $1 \leq i < N$, and presented in a single training pass.

This monotonic progression ensures the model establishes fundamental representations of malicious behavior before encountering more complex samples (i.e., highly obfuscated or heavily contextualized payloads), while avoiding the framework-level instability of discrete stage transitions. The model sequentially optimizes the standard negative log-likelihood objective over this ordered dataset:
\begin{equation}
\mathcal{L}(\theta) = -\frac{1}{N} \sum_{i=1}^N \log p_\theta(y_i|x_i)
\end{equation}

\subsection{DeepSpeed ZeRO-2: Distributed Training}\label{subsec:deepspeed}

Training a Large Language Model (LLM) requires substantial computational resources, often exceeding the memory capacity of a single GPU. While there exist multiple scaling paradigms, they introduce significant trade-offs. Data Parallelism (DP) \citep{li_pytorch_2020} replicates the entire model across all devices, offering no memory reduction per GPU. Alternative strategies, such as Pipeline Parallelism \citep{huang_gpipe_2019} and Model Parallelism \citep{shoeybi_megatron-lm_2020}, distribute memory but often severely compromise compute efficiency, communication overhead, or ease of use.

To overcome these memory limitations, we implemented the Zero Redundancy Optimizer (ZeRO) \citep{rajbhandari_zero_2020} via the DeepSpeed distributed training framework \citep{rasley_deepspeed_2020}. Specifically, we utilized ZeRO Stage 2, which partitions optimizer states and gradients across data-parallel ranks while replicating model parameters:
\begin{equation}
    \text{Memory per GPU} = \Psi_p + \frac{1}{N_d} (\Psi_g + \Psi_{os}) + \Psi_{\text{activation}}
\end{equation}

where $N_d$ is the number of GPUs, $\Psi_p$ represents the replicated model parameters, $\Psi_g$ the gradients, $\Psi_{os}$ the optimizer states, and $\Psi_{\text{activation}}$ the memory required for activations. It is possible to further reduce these GPU memory requirements by integrating CPU offloading via ZeRO-Offload \citep{ren_zero-offload_2021}, which shifts the partitioned optimizer states and gradients directly to host RAM. However, given our hardware setup (4$\times$ A100 80GB GPUs), ZeRO-2 provides an optimal trade-off: because our hardware has sufficient VRAM, we can keep all model weights, gradients, and optimizer states on-device for maximum throughput, avoiding the communication bottlenecks introduced by host-RAM offloading.

\subsection{Anatomy of a Portable Executable (PE) File}\label{subsec:pe_file_composition}

The Portable Executable (PE) format is the standard file format for executables and dynamic-link libraries in 32-bit and 64-bit Windows operating systems, as well as UEFI environments. These files encompass a variety of extensions, most notably \texttt{.exe}, \texttt{.dll}, and \texttt{.sys}. Because our malware analysis focuses exclusively on these file types, an understanding of their composition is essential; it dictates how the OS Loader maps the file into memory and determines where critical behavioral artifacts are stored.

At a high level, a PE file consists of metadata headers followed by a series of distinct data sections. The most critical components for static analysis include:

\begin{itemize}
    \item \textbf{File Headers:} The file begins with the DOS header---identified by the ``MZ'' magic number (4D 5A in hexadecimal)---followed by the PE header. The magic number uniquely distinguishes the file format and serves as a primary identifier. These headers contain essential metadata, such as the target machine architecture, compilation timestamps, and the entry point address.
    
    \item \textbf{Data Sections:} The bulk of the file is divided into sections, each serving a specific purpose and mapped with distinct memory protection attributes (e.g., read, write, execute). By assigning these attributes, the operating system ensures that a section meant only for reading data cannot execute code. The \texttt{.text} section contains the executable machine code, which disassemblers and decompilers translate into Assembly and C code. The \texttt{.data} and \texttt{.rdata} sections store initialized variables, constants, and string literals. Analyzing these sections is crucial for extracting Indicators of Compromise (IoCs) such as embedded IP addresses, URLs, or registry keys.
    
    \item \textbf{Import Address Table (IAT):} To avoid implementing the entirety of the code required to run, PE files rely on external functions provided by the Windows API. The IAT acts as a directory of these external dependencies. Similar to how cryptographic hashes uniquely identify entire files, we can compute a hash of the IAT, known as the \texttt{imphash}. This hash reveals if two files import the exact same libraries and functions in the same order, strongly suggesting they exhibit similar behavior.
\end{itemize}

\textbf{Packed Binary Files} --- To evade detection and hinder static analysis, malware authors frequently utilize packers and obfuscators. These packers compress or encrypt the original executable file, wrapping it in a small decoding stub. Upon execution, this stub runs first, actively unpacking or decrypting the hidden payload directly into memory before transferring the execution flow to it. Consequently, identifying and unpacking these binaries is a mandatory preprocessing step before reliable static feature extraction or decompilation can occur.

\subsection{Graph Representations of Code}\label{subsec:graph_representations}

While code-based representations like assembly and C code provide granular details about a program's instructions, they are highly susceptible to superficial obfuscation techniques such as instruction substitution, dead-code insertion, or register renaming. Graph representations allow analysts to abstract the code into structural topologies. We define these structures formally as directed graphs $G = (V, E)$, where $V$ represents the set of nodes and $E \subseteq V \times V$ represents the directed edges. The two primary structures used are:

\begin{itemize}
    \item \textbf{Control Flow Graph (CFG):} A CFG models the internal execution paths within a single function, defined as $G_{\text{CFG}} = (V_C, E_C)$. The node set $V_C$ consists of basic blocks---sequences of instructions with no internal branching---and the edge set $E_C$ represents the control flow transfers (e.g., jumps, loops, or conditional branches) between them.
    \item \textbf{Function Call Graph (FCG):} An FCG models the macro-level execution flow across the entire program, defined as $G_{\text{FCG}} = (V_F, E_F)$. Here, the node set $V_F$ represents individual functions, and the edge set $E_F$ denotes the calls between them. By mapping this execution flow, an FCG provides a high-level topographical map of the software, revealing structural hubs, central dispatcher functions, and the overall architectural intent of the executable file.
\end{itemize}

By treating code execution as a topological space, these representations bypass noise that would otherwise hinder analysis. To process these structures in machine learning, they must be mapped into compatible modalities. For instance, serializing the graphs into a text-based format, such as the DOT language, provides a textual representation of the topologies. This translation allows the models to ingest complex execution flows through standard contextual memory. Additionally, these structural abstractions can be easily rendered into vector graphics and subsequently into rasterized images, expanding the analysis from purely textual to spatial representations suitable for visual processing techniques.

\subsection{Retrieval-Augmented Generation (RAG)}\label{subsec:rag_theory}

LLMs have a fundamental weakness: they are inherently limited to the data encountered during their training. This data---stored as parametric memory---gets embedded in their weights and serves as their knowledge base. However, due to its static nature, this memory can quickly become stale in rapidly evolving domains or struggle with highly specific topics, often leading to hallucinations. Retrieval-augmented generation (RAG) overcomes this limitation by dynamically fetching relevant external context during inference, grounding the model's responses in verifiable facts and improving domain specificity.

Traditional semantic RAG relies on vector similarity to fetch isolated, descriptive text chunks. Formally, given an input query $x$ and a corpus of external documents $\mathcal{D}$, a retriever encodes both into dense vector representations. It then retrieves a subset of relevant context $\mathcal{Z} \subset \mathcal{D}$ that maximizes a chosen distance metric between the query and document embeddings. The language model then conditions its generation on both the original query and this retrieved context, shifting the standard predictive probability to $P(y|x, \mathcal{Z}) = \prod_i p_\theta(y_i | x, \mathcal{Z}, y_{<i})$. While highly effective for retrieving specific definitions, this isolated chunking can occasionally struggle to resolve multi-hop relational queries. To address the highly interconnected nature of malware ecosystems, GraphRAG structures intelligence as a knowledge graph, defined as $\mathcal{G} = (\mathcal{V}, \mathcal{E})$. In this structure, nodes $v \in \mathcal{V}$ represent distinct entities (e.g., file hashes, IP addresses, API calls), and edges $e \in \mathcal{E}$ represent the semantic relationships between them. Instead of retrieving isolated texts, the system extracts a relevant, connected subgraph $\mathcal{G}_q \subset \mathcal{G}$ based on the query. The model conditions its output on this structural context, $P(y|x, \mathcal{G}_q)$, which preserves semantic relationships and allows the model to traverse complex execution pathways. Utilizing a dual approach maximizes both the precision of isolated fact retrieval and the depth of relational reasoning.

\subsection{Reasoning Augmentation Techniques}\label{subsec:reasoning_augmentation}

Standard instruction-tuning trains LLMs to map inputs directly to final answers through input-response pairs. While efficient, this approach is limited when faced with complex problems that require further investigation before arriving at a conclusion. Consequently, malware analysis requires deep, multi-layered inspection to deduce a program's true intent. Following a multi-step approach helps to reduce logical errors and hallucinations by preventing unfounded answers. To bridge this gap, advanced prompting techniques have been developed to induce deliberate, systematic reasoning within the models.

\begin{itemize}
    \item \textbf{Chain-of-Thought (CoT):} Introduced by \citet{wei_chain--thought_2022}, CoT prompting  addresses the limitations of direct zero-shot reasoning. Instead of immediately generating a final conclusion, CoT induces the model to decompose complex tasks into a sequence of intermediate logical steps. By articulating its reasoning pathway before answering, the model is forced to allocate more computational steps to the problem, effectively preventing premature conclusions and improving accuracy on tasks requiring deep deduction.
    
    \item \textbf{Chain-of-Verification (CoVe):} While CoT improves logical flow, models can still hallucinate intermediate facts. To combat this, \citet{dhuliawala_chain--verification_2024} developed the Chain-of-Verification (CoVe) framework. CoVe requires the model to act as its own auditor through a four-step process: (1) drafting an initial baseline response, (2) formulating independent verification questions to fact-check its own draft, (3) executing those queries against the provided source data, and (4) generating a final, corrected output based on the verified facts. 
\end{itemize}

Integrating these techniques shifts LLMs from simple pattern-matchers to deliberate analytical engines. In the context of malware attribution, where a single misidentified heuristic can derail an entire investigation, utilizing CoT and CoVe ensures the reasoning pipeline remains highly granular, auditable, and factually grounded.

\subsection{Knowledge Distillation in LLMs}\label{subsec:distilled_models}

Deploying LLMs for complex tasks like malware analysis often requires processing massive context windows. Using frontier models for these tasks is computationally prohibitive for widespread use. Knowledge Distillation (KD) provides a theoretical framework to resolve this efficiency bottleneck \citep{hinton_distilling_2015}.

LLM distillation involves transferring the reasoning capabilities and nuanced generation patterns of a massive ``teacher'' model into a much smaller, computationally efficient ``student'' model. This paradigm shifts the training of the model from relying solely on raw data to being trained on the high-quality outputs and reasoning traces generated by the teacher. Consequently, the student model can achieve comparable performance on domain-specific logic tasks while drastically reducing the required GPU memory and inference latency.

\subsection{Continued Pre-Training (CPT) and Supervised Fine-Tuning (SFT)}\label{subsec:cpt_sft}

Adapting an LLM to a highly specialized and syntactically rigid domain, while simultaneously teaching the model to solve specific tasks, requires a structured training pipeline. This adaptation is typically achieved through a two-step framework: Continued Pre-training (CPT) followed by Supervised Fine-Tuning (SFT) \citep{gururangan_dont_2020}.

\begin{itemize}
    \item \textbf{Continued Pre-training (CPT):} Standard LLMs are pre-trained primarily on broad, natural language corpora. Consequently, their internal representations of highly technical or domain-specific languages are often shallow. To resolve this, CPT subjects the model to further self-supervised learning on an unstructured corpus of domain-specific text. For a given sequence $x$, the model optimizes the standard autoregressive next-token prediction objective:
    \begin{equation}
    \mathcal{L}_{\text{CPT}}(\theta) = -\sum_{t=1}^{|x|} \log p_\theta(x_t | x_{<t})
    \end{equation}
    This extended training phase allows the model to map the unique vocabulary, syntax, and implicit structural topologies of the new domain into its parametric memory.
    
    \item \textbf{Supervised Fine-Tuning (SFT):} While CPT focuses on helping the model understand the ``language'' of the target domain, SFT aligns that foundational knowledge to perform specific analytical tasks. To achieve this, the model is trained on a curated dataset of structured input-response pairs. For a given input $x$ and target response $y$, the loss is calculated conditionally, measuring the likelihood of the response given the prompt:
    \begin{equation}
    \mathcal{L}_{\text{SFT}}(\theta) = -\sum_{t=1}^{|y|} \log p_\theta(y_t | x, y_{<t})
    \end{equation}
    This phase teaches the model how to adhere to human instructions, format its output, and apply its domain knowledge to execute specific reasoning pathways.
\end{itemize}

By decoupling domain immersion (CPT) from task alignment (SFT), this two-stage framework prevents the model from simply memorizing formatting templates, ensuring it develops a robust, generalizable understanding of the underlying domain before being constrained to specific instructional tasks.

\subsection{Parameter-Efficient and Quantized Fine-Tuning}\label{subsec:peft_theory}

Full fine-tuning of LLMs is computationally prohibitive on standard hardware, as it requires updating billions of parameters. Parameter-Efficient Fine-Tuning (PEFT) methods mitigate this by freezing the pre-trained weights and training a minimal set of auxiliary parameters \citep{houlsby_parameter-efficient_2019}.

\paragraph{Low-Rank Adaptation (LoRA).}
Low-Rank Adaptation (LoRA) \citep{hu_lora_2021} is a foundational PEFT architecture. Instead of computing a full-rank weight update $\Delta W \in \mathbb{R}^{d \times k}$, LoRA freezes the pre-trained model weights $W_0 \in \mathbb{R}^{d \times k}$ and injects trainable low-rank decomposition matrices, $B \in \mathbb{R}^{d \times r}$ and $A \in \mathbb{R}^{r \times k}$, into each target layer. The forward pass is modified such that the frozen pre-trained weights and the low-rank adapter matrices are computed in parallel:
\begin{equation}
    W = W_0 + \frac{\alpha}{r} BA
\end{equation}
where $r \ll \min(d, k)$ is the rank and $\alpha$ is a scaling constant. Because the inner dimension $r$ is strictly limited, LoRA drastically reduces the number of trainable parameters and memory footprint while maintaining competitive performance.

\paragraph{Quantization-Aware Adaptation (QLoRA).}
While LoRA reduces trainable parameters, the base model still occupies significant VRAM. QLoRA \citep{dettmers_qlora_2023} integrates high-precision quantization into the PEFT paradigm by quantizing the frozen backbone to 4-bit NormalFloat (NF4) precision. QLoRA further employs double quantization, which quantizes the quantization constants themselves, reducing their memory footprint without degrading numerical stability. During the forward and backward passes, quantized weights are dequantized on-the-fly to compute activations.

\paragraph{Rank-Stabilized LoRA (rsLoRA).}
Standard LoRA scales adapter activations by a factor of $\alpha/r$. Consequently, as the rank $r$ increases to capture more complex domain knowledge, the learning slows down and performance plateaus due to vanishing update magnitudes. Rank-Stabilized LoRA (rsLoRA) \citep{kalajdzievski_rank_2023} addresses this by replacing the scaling factor with $\alpha/\sqrt{r}$. This correction allows the model to utilize higher ranks without gradient instability or the need for extensive learning rate tuning.

\paragraph{Weight-Decomposed LoRA (DoRA).}
While standard LoRA alters both the magnitude and direction of the model's weight vectors simultaneously, Weight-Decomposed LoRA (DoRA) \citep{liu_dora_2024} decomposes the pre-trained weight matrix $W_0$ into independent magnitude $m$ and directional $V$ components, such that $W_0 = m \frac{V}{\|V\|_c}$, where $\|\cdot\|_c$ denotes the vector-wise norm. By freezing the pre-trained weights and applying the low-rank update exclusively to the directional component, the adapted weight matrix becomes:
\begin{equation}
    W' = m \frac{W_0 + BA}{\|W_0 + BA\|_c}
\end{equation}
By decoupling these updates, DoRA closely mimics the learning capacity and update patterns of full fine-tuning while retaining the parameter efficiency of standard LoRA.

%~~~~~~~~~~~~~~~~~~~~~~~~~~~~~~~~~~~~~~~~~~~~~~~~~~~~~~~~~~~~~~~~~~~~~~~~~~~~~~~~~~~~~~~~
\section{Related Work}\label{sec:related_work}
Malware analysis is commonly divided into static and dynamic analysis. Static analysis examines a sample without execution, using artifacts such as PE headers, imported APIs, assembly, decompiled code, CFGs, and FCGs, while dynamic analysis observes runtime behavior in a controlled environment. This research focuses on static Windows PE malware analysis for early, code-centric malware attribution and analyst-oriented reasoning.
\subsection{Existing Malware Datasets}\label{subsec:existing_datasets}

\begin{table*}[h]
\centering
\scriptsize
\caption{Availability of related PE malware datasets for code-centric malware attribution and LLM-oriented analysis tasks. Dec. Code = Decompiled Code, CFG = Control Flow Graph, FCG = Function Call Graph, Mal. Code Seg. = Malicious Code Segment, Vuln. Code/CWE = Vulnerable Code/Common Weakness Enumeration, API = Application Programming Interface, CTI = Cyber Threat Intelligence, IoC = Indicator of Compromise, LLM = Large Language Model;  \\ \protect\customcirc{emptycirc} = ``Not Available'', \protect\customcirc{halfcirc} = ``Partially Available'', and \protect\customcirc{fullcirc} = ``Available''.}
\label{tab:lccd_dataset_comparison}
\scriptsize
\renewcommand{\arraystretch}{1.35}
\setlength{\tabcolsep}{4pt}
\resizebox{\linewidth}{!}{%
\begin{tabular*}{\linewidth}{@{\extracolsep{\fill}}lccccccccc}
\toprule
\textbf{Dataset} &
\textbf{Dec. Code} &
\textbf{CFG/FCG} &
\textbf{Mal. Code Seg.} &
\textbf{Vuln. Code/CWE} &
\textbf{API Risk} &
\textbf{CTI/IoC} &
\textbf{Attribution} &
\textbf{LLM Data} &
\textbf{Analyst Tasks} \\
\midrule

Malicia \cite{nappa2015malicia} &
\customcirc{emptycirc} &
\customcirc{emptycirc} &
\customcirc{emptycirc} &
\customcirc{emptycirc} &
\customcirc{emptycirc} &
\customcirc{emptycirc} &
\customcirc{fullcirc} &
\customcirc{emptycirc} &
\customcirc{emptycirc} \\

Microsoft BIG-2015\cite{ronen2018microsoft} &
\customcirc{emptycirc} &
\customcirc{emptycirc} &
\customcirc{emptycirc} &
\customcirc{emptycirc} &
\customcirc{emptycirc} &
\customcirc{emptycirc} &
\customcirc{fullcirc} &
\customcirc{emptycirc} &
\customcirc{emptycirc} \\

EMBER 2018 \cite{anderson2018ember} &
\customcirc{emptycirc} &
\customcirc{emptycirc} &
\customcirc{emptycirc} &
\customcirc{emptycirc} &
\customcirc{emptycirc} &
\customcirc{emptycirc} &
\customcirc{emptycirc} &
\customcirc{emptycirc} &
\customcirc{emptycirc} \\

SOREL-20M \cite{harang2020sorel} &
\customcirc{emptycirc} &
\customcirc{emptycirc} &
\customcirc{emptycirc} &
\customcirc{emptycirc} &
\customcirc{emptycirc} &
\customcirc{emptycirc} &
\customcirc{emptycirc} &
\customcirc{emptycirc} &
\customcirc{emptycirc} \\

BODMAS \cite{yang2021bodmas} &
\customcirc{emptycirc} &
\customcirc{emptycirc} &
\customcirc{emptycirc} &
\customcirc{emptycirc} &
\customcirc{emptycirc} &
\customcirc{emptycirc} &
\customcirc{fullcirc} &
\customcirc{emptycirc} &
\customcirc{emptycirc} \\

CIC-SGG-2024 \cite{mohammadian2025explainable} &
\customcirc{emptycirc} &
\customcirc{fullcirc} &
\customcirc{emptycirc} &
\customcirc{emptycirc} &
\customcirc{emptycirc} &
\customcirc{emptycirc} &
\customcirc{fullcirc} &
\customcirc{emptycirc} &
\customcirc{emptycirc} \\

FCG-MFD \cite{hadi2025fcg} &
\customcirc{emptycirc} &
\customcirc{fullcirc} &
\customcirc{emptycirc} &
\customcirc{emptycirc} &
\customcirc{emptycirc} &
\customcirc{emptycirc} &
\customcirc{fullcirc} &
\customcirc{emptycirc} &
\customcirc{emptycirc} \\

SBAN \cite{jelodar2025sban} &
\customcirc{fullcirc} &
\customcirc{emptycirc} &
\customcirc{emptycirc} &
\customcirc{emptycirc} &
\customcirc{emptycirc} &
\customcirc{emptycirc} &
\customcirc{fullcirc} &
\customcirc{halfcirc} &
\customcirc{emptycirc} \\

\textbf{LCCD (Ours)} &
\customcirc{fullcirc} &
\customcirc{fullcirc} &
\customcirc{fullcirc} &
\customcirc{fullcirc} &
\customcirc{fullcirc} &
\customcirc{fullcirc} &
\customcirc{fullcirc} &
\customcirc{fullcirc} &
\customcirc{fullcirc} \\

\bottomrule
\end{tabular*}%
}
\end{table*}

Several malware datasets have been developed to support malware detection, classification, and family identification. Early datasets such as Malicia~\cite{nappa2015malicia} and Microsoft BIG-2015~\cite{ronen2018microsoft} mainly focus on malware family classification using binaries, byte sequences, or family labels. Similarly, EMBER~\cite{anderson2018ember}, SOREL-20M~\cite{harang2020sorel}, and BODMAS~\cite{yang2021bodmas} provide large-scale Windows PE samples, static features, metadata, and labels for malware detection and classification. These datasets have significantly advanced static malware analysis; however, most of them are designed for feature-based machine learning rather than code-centric reasoning. As shown in~\cref{tab:lccd_dataset_comparison}, they generally do not provide decompiled code, malicious code segment annotations, vulnerable code segments, suspicious API risk context, CTI/IoC enrichment, or LLM-oriented instruction data.

Recent datasets have started to provide richer representations for malware analysis. For example, CIC-SGG-2024~\cite{mohammadian2025explainable} introduces graph-based malware representations using Control Flow Graphs (CFGs) and Function Call Graphs (FCGs), making it useful for structural malware analysis. FCG-MFD~\cite{hadi2025fcg} extends malware benchmarking through a large and modern classifier-evaluation setting, while SBAN~\cite{jelodar2025sban} moves toward multimodal and LLM-oriented software code mining. Although these datasets broaden the malware dataset landscape, they still do not jointly support malware attribution, malicious code segment identification, vulnerable code/CWE analysis, suspicious API interpretation, MITRE ATT\&CK mapping, CTI/IoC grounding, and analyst-oriented LLM tasks.

Furthermore, the proposed LCCD is specifically designed to overcome the limitations of existing malware datasets for LLM-based malware attribution. Existing datasets typically support either feature-based detection, family classification, API-call analysis, or graph-based modeling, but they do not preserve the complete evidence chain required for analyst-oriented malware reasoning. In contrast, LCCD links each Windows PE sample with complementary reverse-engineering artifacts, including decompiled C code, assembly code, CFG/FCG structures, PE metadata, suspicious API evidence, CTI/IoC context, malicious code segment information, vulnerable code/CWE evidence, and vector embeddings. This design enables LLMs to move beyond label prediction and reason over the code, structure, behavior, and threat-intelligence evidence supporting an attribution decision. To the best of our knowledge, LCCD is the first dataset that jointly provides code-centric malware artifacts and structured instruction-tuning data for multi-task malware attribution, including malicious code analysis, vulnerability analysis, API behavior reasoning, MITRE ATT\&CK mapping, detection guidance, and analyst-ready report generation.

\subsection{Non-LLM-Based Malware Detection} \label{subsec:non_llm_detection}
Before the adoption of LLMs, malware detection and family classification were mainly addressed using signature-based methods \cite{fortino2023sigil}, handcrafted static features \cite{jeon2024static}, machine learning \cite{gardiner2016security, ucci2019survey}, deep learning \cite{gopinath2023comprehensive}, and graph-based representations \cite{hadi2025fcg}. In static analysis, researchers commonly extract byte sequences, opcode patterns, imported APIs, PE header fields, entropy values, strings, CFGs, and FCGs to train classifiers for malware detection or family attribution~\cite{b31,b33,b34,b38,b39,b40}. Moreover, image-based approaches convert binaries into grayscale images and use attention mechanisms to identify discriminative regions for malware family classification~\cite{b31}. In addition, graph-based approaches model program structure through CFGs and FCGs; for example, CFG Explainer identifies influential CFG nodes or subgraphs for GNN-based malware classification~\cite{b32,tang2026hhgdroid}. Although these methods have improved malware detection and classification, they remain largely optimized for prediction and often provide limited analyst-oriented evidence, such as malicious code segment localization and vulnerable code reasoning.

Dynamic analysis, on the other hand, executes malware in a controlled environment to observe runtime behaviors such as process creation, file-system activity, registry modification, memory behavior, and network communication~\cite{b35,b36}. Although this approach provides valuable behavioral evidence, it can be costly, time-consuming, and difficult to scale for large malware collections. Moreover, advanced malware may use evasion techniques, including anti-sandboxing, delayed execution, packing, and environment fingerprinting, to hide malicious behavior during analysis~\cite{b37}. Consequently, static analysis remains important for early malware triage and large-scale classification~\cite{b38,b39,b40}. However, non-LLM-based methods often remain sensitive to feature drift, obfuscation, packing, code reuse, and changes in malware family behavior. As a result, they may struggle to identify previously unseen variants, zero-day malware, or samples whose observable features have been modified to evade known detection patterns. This limitation motivates code-centric LLM-assisted frameworks that can combine reverse-engineering artifacts and analyst-oriented reasoning.

\subsection{LLM-Based Malware Analysis and Cybersecurity Applications}
Recent studies have increasingly explored LLMs for malware detection, reverse engineering, forensic triage, and cyber threat intelligence. Jelodar et al.~\cite{jelodar2025large} reviewed LLM applications in malware code detection, generation, monitoring, reverse engineering, and family analysis, highlighting the growing role of LLMs in understanding malicious code semantics. However, their work is survey-oriented and does not introduce a deployable malware attribution framework or a code-centric benchmark dataset. In addition, several recent studies have focused on Android malware analysis \cite{zhao2025apppoet}. Feng et al.~\cite{feng2025llm} proposed LLM-MalDetect, which uses string-based APK features, prompt engineering, and LLM fine-tuning for Android malware detection. Similarly, Priambodo et al.~\cite{priambodo2025malqwen} introduced MalQwen, a LoRA fine-tuned Qwen model for generating Android malware analysis reports from decompiled code and expert-labeled reports. Although these studies demonstrate the usefulness of LLMs for malware detection and report generation, they remain Android-specific and do not address Windows PE malware attribution.

Furthermore, for Windows PE malware, Marais et al.~\cite{marais2025semantic} proposed a semantic preprocessing approach for LLM-based malware analysis. Their method converts PE files into analyst-readable JSON reports. Additionally, these reports are enriched with static and behavioral features, packer signatures, YARA matches, MITRE ATT\&CK mappings, and Malware Behavior Catalog knowledge. This representation improves interpretability for malware classification; however, the work mainly focuses on malware category classification and does not provide a multimodal code-centric dataset or an instruction-tuning corpus for attribution-oriented LLM reasoning. Furthermore, recent studies have examined the robustness, classification reliability, and reverse-engineering capabilities of LLMs in malware analysis. Böke and Torka~\cite{boke2025digital} evaluated LLM-based malware analysis under LLVM-level obfuscation and showed that model performance can degrade under compiler-level transformations, highlighting the need for stronger grounding in low-level reverse-engineering evidence. Similarly, Bai et al.~\cite{bai2026automated} proposed a weighted hierarchical ensemble of LLMs for zero-label malware family classification, improving robustness through decision-level aggregation. However, the output remains primarily a family label rather than a complete analyst-facing evidence chain. In addition, Jelodar et al.~\cite{jelodar2026llm4codere} introduced LLM4CodeRE for bidirectional assembly-to-source and source-to-assembly translation, demonstrating the value of LLMs for malware-aware reverse engineering. Nevertheless, its focus is decompilation and code translation rather than malware attribution, IoC validation, MITRE mapping, or threat-intelligence-grounded analysis.

Overall, prior LLM-based malware studies have advanced Android detection, PE semantic preprocessing, malware report generation, family classification, reverse engineering, memory forensics, and edge deployment. However, most focus on detection accuracy, report generation, family-label prediction, forensic summarization, or decompilation. They do not jointly provide a Windows PE-focused benchmark dataset and an evidence-grounded LLM framework. LCC-LLM addresses this gap by combining a code-centric dataset with an expert-system framework for malware attribution, IoC validation, MITRE ATT\&CK mapping, vulnerability analysis, containment guidance, and structured malware reporting.

%~~~~~~~~~~~~~~~~~~~~~~~~~~~~~~~~~~~~~~~~~~~~~~~~~~~~~~~~~~~~~~~~~~~~~~~~~~~~~~~~~~~~~~~~
\section{Methodology for Creating the Large-Scale Code-Centric Dataset}\label{sec:methodology_dataset}

%%%%%%%%%%%%%%%%%%%%%%%%%%%%%%%%%%%%%%%%%%%%%%%%%%%%%%%%%%%%%%%
% FIGURE: DATASET GENERATION PIPELINE
%%%%%%%%%%%%%%%%%%%%%%%%%%%%%%%%%%%%%%%%%%%%%%%%%%%%%%%%%%%%%%%
\begin{figure*}[pos=htbp]
    \centering
    \includegraphics[width=\textwidth]{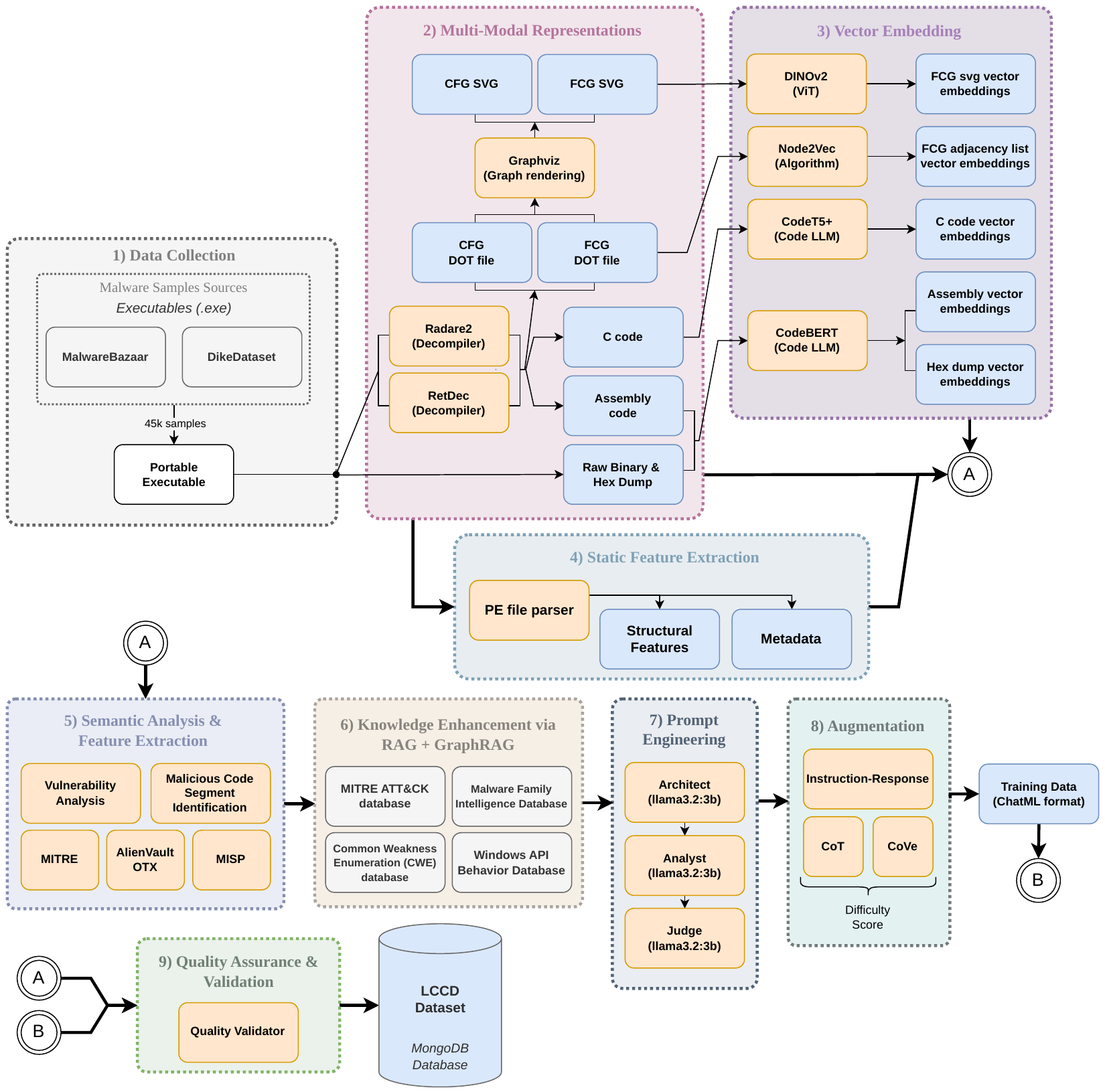}
    \caption{High-level overview of the LCCD generation pipeline. The multi-stage process spans from raw binary collection (1) through multimodal feature extraction and semantic analysis (2-6), to prompt generation and augmentation (7-8), before final database ingestion (9).}
    \label{fig:lccd_pipeline}
\end{figure*}
%==============================================================

%%%%%%%%%%%%%%%%%%%%%%%%%%%%%%%%%%%%%%%%%%%%%%%%%%%%%%%%%%%%%%%
% FIGURE: SAMPLE ANATOMY
%%%%%%%%%%%%%%%%%%%%%%%%%%%%%%%%%%%%%%%%%%%%%%%%%%%%%%%%%%%%%%%

%==============================================================

%%%%%%%%%%%%%%%%%%%%%%%%%%%%%%%%%%%%%%%%%%%%%%%%%%%%%%%%%%%%%%%
% FIGURE: RETDEC PIPELINE
%%%%%%%%%%%%%%%%%%%%%%%%%%%%%%%%%%%%%%%%%%%%%%%%%%%%%%%%%%%%%%%
\begin{figure}[pos=htbp]
    \centering
    \includegraphics[width=\columnwidth]{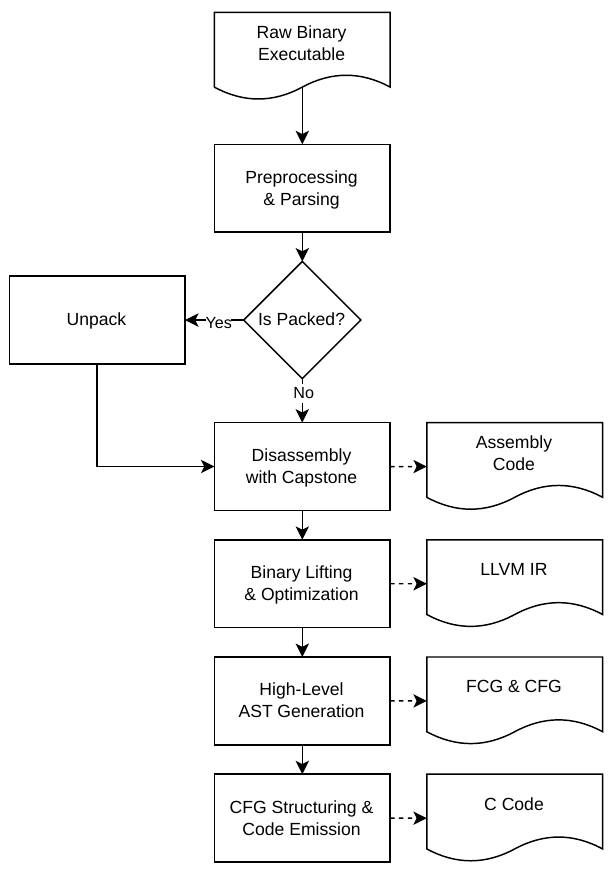}
    \caption{The decompilation pipeline architecture of RetDec.}
    \label{fig:retdec_pipeline}
\end{figure}
%==============================================================

\begin{figure*}[pos=htbp]
    \centering
    \includegraphics[width=0.80\textwidth]{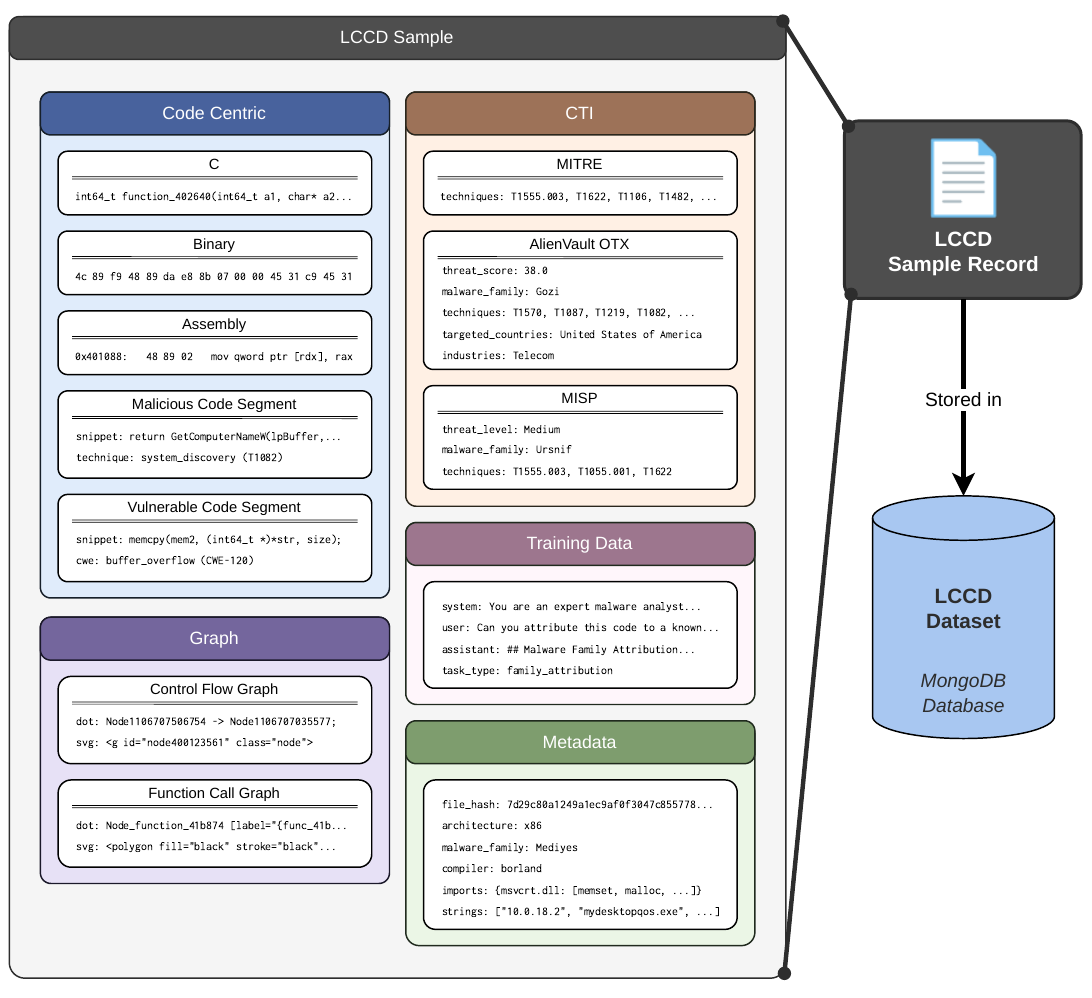}
    \caption{Anatomy of a finalized LCCD sample record. Each structured entry aggregates code-centric representations, topological graphs, enriched CTI metadata, and the generated instruction-tuning prompt configurations.}
    \label{fig:sample_anatomy}
\end{figure*}

We developed LCCD for instruction-tuning Large Language Models (LLMs) on code-centric tasks, specifically focusing on malware family attribution and malicious code segment identification. The methodology relies on a multi-stage pipeline that integrates data collection, multimodal feature extraction, vector embedding, semantic analysis, and prompt engineering augmented by Cyber Threat Intelligence (CTI). \cref{fig:lccd_pipeline} illustrates the high-level architecture of this pipeline, tracing the lifecycle of a malware sample from initial ingestion to final quality validation. Through this process, raw binaries are transformed into comprehensive, structured records. The anatomy of these finalized LCCD samples is visually detailed in~\cref{fig:sample_anatomy}, while the step-by-step logic driving this transformation is formalized in~\cref{alg:lccd_gen}.

%%%%%%%%%%%%%%%%%%%%%%%%%%%%%%%%%%%%%%%%%%%%%%%%%%%%%%%%%%%%%%%
% ALGORITHM: DATASET GENERATION PIPELINE
%%%%%%%%%%%%%%%%%%%%%%%%%%%%%%%%%%%%%%%%%%%%%%%%%%%%%%%%%%%%%%%
\begin{algorithm}
\caption{LCCD Generation Pipeline}
\scriptsize
\label{alg:lccd_gen}
\begin{algorithmic}[1]
    \Require DikeDataset $\mathcal{D}_{Dike}$, MalwareBazaar $\mathcal{D}_{MB}$, CTI $\mathcal{K}_{CTI}$
    \Ensure Unified Dataset $\mathcal{D}_{LCCD}$ (Multimodal, Features, \& Instruction-Tuned Prompts)
    
    \State \textbf{// Phase 1: Data Collection \& Preprocessing}
    \State $\mathcal{S} \gets \text{FetchAndFilter}(\mathcal{D}_{Dike} \cup \mathcal{D}_{MB})$ \label{lccd:line:data_fetch}
    
    \For{each malware sample $s \in \mathcal{S}$}
        \State \textbf{// Phase 2: Multimodal Representations}
        \State $M_s \gets \text{ExtractViews}(s)$ \label{lccd:line:extract_views}
        
        \State \textbf{// Phase 3: Vector Embedding}
        \For{overlapping window $w$ in $M_s(\text{C}, \text{Asm}, \text{Hex})$} \label{lccd:line:embedding_start}
            \State Add $\text{Embed}(w)$ to $E_s$
        \EndFor
        \State Add $\text{Embed}(M_s(\text{Graphs, SVGs}))$ to $E_s$ \label{lccd:line:embedding_end}
    
        \State \textbf{// Phase 4: Static Analysis \& Labeling}
        \State $f_{static} \gets \text{ExtractStaticFeatures}(M_s)$ \label{lccd:line:static_features}
        \State $f_{sem}, \text{score}_{diff} \gets \text{MapToTaxonomies}(M_s, f_{static})$ \label{lccd:line:map_taxonomies}
        
        \State $L_{rep} \gets \text{QueryCTI}(\mathcal{K}_{CTI}, s)$ \label{lccd:line:query_cti}
        \State $L_{imp} \gets \text{CrossReferenceBy}(\text{imphash}, s)$ \label{lccd:line:reference_imphash}
        \If{$(L_{rep} \cup L_{imp}) \neq \emptyset$}
            \State $\text{label} \gets \text{StandardizeWithAVClass}(L_{rep} \cup L_{imp})$ \label{lccd:line:standardize_avclass}
        \Else
            \State $\text{label} \gets \text{``Unknown''}$ \label{lccd:line:label_unknown}
        \EndIf
    
        \State \textbf{// Phase 5: RAG \& Prompt Engineering}
        \State $c_{rag} \gets \text{RetrieveRAGContext}(\mathcal{K}_{CTI}, f_{static}, f_{sem})$ \label{lccd:line:rag_context}
        \State $p_{plan} \gets \text{Architect}(M_s, c_{rag})$ \label{lccd:line:prompt_architect} 
        \State $r_{raw} \gets \text{Analyst}(p_{plan})$ \label{lccd:line:prompt_analyst}
        \State $r_{fin} \gets \text{Judge}(r_{raw}, p_{plan}, \text{ground\_truth})$ \label{lccd:line:prompt_judge}
        
        \State Add $(s, M_s, E_s, f_{static}, f_{sem}, \text{label}, r_{fin}, \text{score}_{diff})$ to $\mathcal{D}_{LCCD}$ \label{lccd:line:aggregation}
    \EndFor
    
    \State \textbf{// Phase 6: Quality Assurance}
    \State $\mathcal{D}_{LCCD} \gets \text{ValidateAndBackfill}(\mathcal{D}_{LCCD})$ \label{lccd:line:validate}
    
    \State \Return $\mathcal{D}_{LCCD}$
\end{algorithmic}
\end{algorithm}
%==============================================================

\subsection{Data Collection}\label{subsec:data_collection}
We collected malware samples from the DikeDataset~\cite{iosif_dikedataset_2021} and MalwareBazaar~\cite{abusech_malwarebazaar_nodate}, focusing specifically on Windows Portable Executable (PE) files, as shown in~\cref{lccd:line:data_fetch}. To ensure temporal relevance and capture recent or ongoing malware campaigns, the MalwareBazaar collection was limited to samples submitted between January 2020 and January 2026. This resulted in a total of 35,000 samples, each uniquely identified by its SHA-256 hash to prevent duplicates and streamline CTI integration. MalwareBazaar provides two methods to fetch malware samples: an API and a data lake that gets updated with batch archives of samples divided by day and month. Due to the vast volume of samples we retrieved, we used the data lake archives to avoid overwhelming MalwareBazaar’s API.

%%%%%%%%%%%%%%%%%%%%%%%%%%%%%%%%%%%%%%%%%%%%%%%%%%%%%%%%%%%%%%%
% FIGURE: SAMPLE ANATOMY
%%%%%%%%%%%%%%%%%%%%%%%%%%%%%%%%%%%%%%%%%%%%%%%%%%%%%%%%%%%%%%%
%==============================================================

Both sources contain a variety of file formats in addition to PE files, so we filtered to only include in the dataset the file extension .exe, even though this format can have multiple extensions like .dll, .sys, .efi. In addition to the file format filtering, we performed an initial precheck of the file headers to remove samples that were broken. We performed this check with the intention of increasing the percentage of successful decompilations. We separated the collected files by size to have a greater diversity among samples: Small ($\le$~100KB), Medium ($\le$~500KB), and Large ($\le$~5MB). The smaller samples dominate our dataset with 22,106, followed by the medium ones with 12,192, and lastly the large ones with 394, thus having a ratio of 55:30:1. We decided to include samples following this ratio because of the longer processing time the larger samples take.

\subsection{Multimodal Representations}\label{subsec:multimodal_representations}

We processed the malware samples into different code and structural representations to capture their multifaceted nature (\cref{lccd:line:extract_views}). Additionally, each representation may capture aspects of the malware that are lost in another one, so analyzing them all enriches the analysis. The transformation is achieved by implementing a unified decompilation pipeline using RetDec~\cite{avast_software_retdec_2026} as the primary source and Radare2~\cite{radare_org_radare2_2026} as fallback and enrichment. RetDec first preprocesses the binary data to identify the file sections, extract debugging information, and unpack it. Following this step, it utilizes the Capstone Engine~\cite{anh_quynh_capstone_2026} to disassemble the binary data into assembly code. These assembly instructions are then lifted into LLVM intermediate representation (IR) and optimized, simplifying and reconstructing the logic. RetDec proceeds to translate the LLVM IR into a back-end IR that serves as a high-level abstract syntax tree (AST). This representation helps reconstruct structured loops and conditionals, essential for finally generating the C code. RetDec's pipeline to convert the binary data into decompiled C code is depicted in~\cref{fig:retdec_pipeline}. While obfuscation often causes decompilation to fail, our dual-engine approach maximizes data recovery by merging outputs from both tools. Furthermore, Radare2’s native support for legacy and obscure formats ensures successful decompilation where other tools typically fail. With the previous unified decompilation pipeline, we obtained the C code, assembly code, FCG, and CFG. Additionally, we dumped the binary data as hexadecimal (hex) so it can be ingested by an LLM as text. The structural representations were further processed to obtain a visual representation of the topologies as vector images. We did this last conversion using Graphviz due to its native handling of the DOT files that are generated by RetDec, tool maturity, and easy headless integration into our pipeline.

\subsection{Vector Embedding Strategy}\label{subsec:vector_embedding}

We then precomputed the vector embeddings for each of the multimodal representations (\crefrange{lccd:line:embedding_start}{lccd:line:embedding_end}). While these embeddings will not be used as input for fine-tuning an LLM, they serve as fingerprints for the data. These fingerprints can be used for sorting, searching, cleaning, and visualizing the data in the dataset. For instance, they can enable semantic search in RAG pipelines, as demonstrated by \citet{zhang_tracerag_2025} with TraceRAG, or be used to filter duplicates and retain only high-quality samples, akin to SemDeDup’s semantic deduplication approach of \citet{abbas_semdedup_2023}.

We utilized two distinct models to generate embeddings for the multimodal code representations: CodeT5+~\cite{wang_codet5_2023} and CodeBERT~\cite{feng_codebert_2020}. CodeT5+ was selected to embed the decompiled C code as it is optimized for understanding programming semantics. This model was pre-trained with bimodal text-code data from 8 programming languages, including C. Conversely, we embedded the assembly code and hex dump using CodeBERT. Even though the model was trained on high-level languages, \citet{yuki_asmdocgen_2026} demonstrated in AsmDocGen the model’s ability to learn assembly code. Their work suggests that CodeBERT's capacity to capture syntactic and semantic information translates effectively to understanding the unique grammar of low-level code. Extending this rationale, we leverage the model’s self-attention mechanism to capture the patterns and structural dependencies within the hex dump; therefore, treating the byte sequences as a low-level architectural representation of the program logic.

These code representations of malware tend to exceed the context window of the embedding models, necessitating a strategy to handle these long sequences. Because it is not possible to process the entire code logic in a single pass, we implemented a 512-token sliding window strategy with a 256-token stride. These overlapping segments ensure that long-range dependencies and semantic indicators are captured even if they span across distinct fragments. The embeddings from all the individual chunks are ultimately aggregated via mean pooling to produce the final 768-dimensional vector for each representation.

Furthermore, we employed a bimodal strategy that captures both the topological structures for relational data and the spatial features for visual analysis. We mapped the topology of the FCG with Node2Vec~\cite{grover_node2vec_2016}, a technique that performs random walks across a graph to identify local neighborhoods and structural hubs among functions. To capture these structural dependencies effectively, we configured the model to generate 128-dimensional node embeddings. We simulated 200 random walks per node, with each walk spanning a length of 30 nodes. Additionally, we set up both the return parameter ($p$) and the in-out parameter ($q$) to 1.0, making the walks remain unbiased. This balances the exploration of local and tightly connected function clusters with the discovery of broader structural roles. Complementing this strategy, we converted the FCGs into rasterized images to treat the code execution as a visual topology problem and harness established vision techniques. While rasterizing vector graphs inherently causes some information loss, such as dropping explicit node labels and making individual connections difficult to discern in overly dense networks, it enables us to observe macro-level topological shapes that bypass standard textual obfuscation. These visual footprints can reveal distinct structural signatures, such as prominent hubs or recurring geometric formations, that are indicative of specific malware families or categories. We generated 768-dimensional vector embeddings from these images using DINOv2~\cite{oquab_dinov2_2024}, a self-supervised vision model highly adept at recognizing complex structural shapes and visual patterns.

\subsection{Threat Intelligence Integration and Profiling}\label{subsec:threat_intelligence}

Following decompilation, we performed static analysis on the C and assembly code to extract behavioral artifacts without executing the binaries (\cref{lccd:line:static_features}). By examining these outputs, we isolated critical features such as imported libraries and function dependencies. Additionally, we mined the code for embedded string literals to uncover IoCs. These artifacts include API names, Command-and-Control (C2) IP addresses and domains, and targeted file paths. This information provides context for understanding the sample’s intended execution flow, serving as key telemetry for threat actor attribution.

We analyzed the static features and high-level decompiled C code using a combination of heuristics and standard malware taxonomies from CTI frameworks (\cref{lccd:line:query_cti}). Specifically, we identified vulnerabilities in the decompiled C code by matching syntactic patterns to those in the Common Weakness Enumeration (CWE) database. Likewise, we precisely identified malicious code segments and their parent functions using a curated database of indicators. Segments exhibiting malicious behaviors were then mapped to MITRE ATT\&CK techniques, creating a comprehensive behavior profile for each sample.

%%%%%%%%%%%%%%%%%%%%%%%%%%%%%%%%%%%%%%%%%%%%%%%%%%%%%%%%%%%%%%%
% FIGURE: DIFFICULTY SCORING
%%%%%%%%%%%%%%%%%%%%%%%%%%%%%%%%%%%%%%%%%%%%%%%%%%%%%%%%%%%%%%%
\begin{figure}[pos=htbp]
    \centering
    \includegraphics[width=\columnwidth]{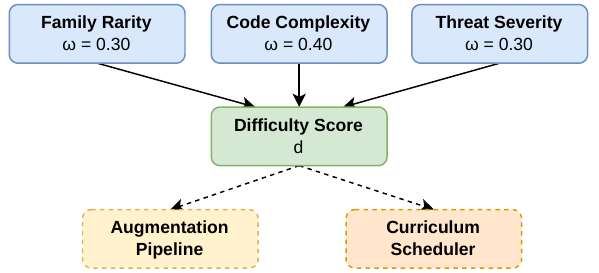} 
    \caption{Difficulty scoring model. The score determines both the augmentation mode applied to each sample and its position in the curriculum-ordered dataset.}
    \label{fig:difficulty_score}
\end{figure}
%==============================================================

Using this aggregate of extracted features, we implemented a multidimensional scoring system to assess the complexity of each sample, as depicted in~\cref{fig:difficulty_score}. By calculating a weighted score that considers the decompiled C code, imports, implemented techniques, malware family rarity, severity of the threat, and level of obfuscation, we assigned each sample both a numeric value and a discrete category. This classification provides a clear progression from trivial to expert-level samples, categorizing them by the inherent difficulty their internal characteristics pose during analysis.

\subsection{Automated Labeling and Categorization Pipeline}\label{subsec:labeling}

While the samples sourced from the DikeDataset were already labeled, we required a strategy to systematically classify the unlabeled PE files retrieved from MalwareBazaar. For the DikeDataset subset, we opted to utilize the ground truth from the ``Malware Detection PE-Based Analysis Using Deep Learning Algorithm Dataset'' (malpe\_dp)~\cite{anh_pham_tuan_malware_2018}, which served as one of DikeDataset's foundational sources. This substitution was necessary because the original DikeDataset labels only provided broad malware categories rather than specific family names.

%%%%%%%%%%%%%%%%%%%%%%%%%%%%%%%%%%%%%%%%%%%%%%%%%%%%%%%%%%%%%%%
% FIGURE: LABELING PIPELINE
%%%%%%%%%%%%%%%%%%%%%%%%%%%%%%%%%%%%%%%%%%%%%%%%%%%%%%%%%%%%%%%
\begin{figure}[pos=htbp]
    \centering
    \includegraphics[width=\columnwidth]{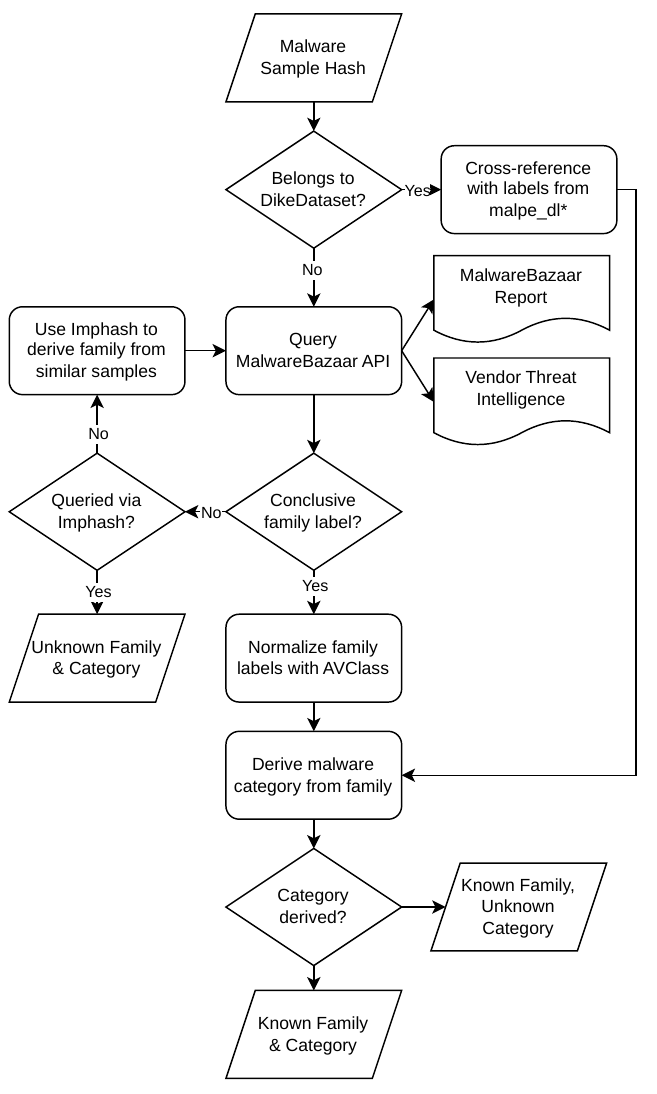} 
    \caption{Flowchart of the malware labeling pipeline. The process determines the family and category of a sample by first checking the local DikeDataset, and subsequently querying the MalwareBazaar API with fallback Imphash lookups for unknown samples. Labels are normalized via AVClass prior to final categorization. (*malpe\_dl: Malware Detection PE-Based Analysis Using Deep Learning Algorithm Dataset).}
    \label{fig:labeling_pipeline}
\end{figure}
%==============================================================

For the remaining samples, we implemented the labeling pipeline depicted in~\cref{fig:labeling_pipeline}. We initially queried the MalwareBazaar API using the SHA-256 hash of each sample to retrieve its associated threat report (\cref{lccd:line:query_cti}). Because these reports do not always provide a conclusive malware family, we corroborated the data with secondary Cyber Threat Intelligence (CTI) sources linked within the reports. This involved analyzing telemetry from dynamic analysis sandboxes, such as ANY.RUN and CAPE Sandbox, which are designed to capture complex behavioral signatures.

When direct attribution from these reports was inconclusive, we leveraged the import hash (imphash) to identify structural similarities with known samples (\cref{lccd:line:reference_imphash}). Unlike cryptographic hashes (e.g., SHA-256) that change entirely with single-bit modifications, an imphash remains consistent across binaries that import the exact same libraries and API functions. While shared imports are not definitive proof of identical behavior, the imphash served as a reliable fallback heuristic for family attribution when all other strategies were exhausted.

Once the initial family labels were attributed, we normalized the outputs using AVClass~\cite{malicia_lab_avclass_2023}, an open-source Python tool designed to standardize malware family names across varying antivirus vendors (\cref{lccd:line:standardize_avclass}). Following normalization, we mapped 2,226 distinct family names to their broader malware categories. To achieve this efficiently, we utilized Gemini 3.1 Pro to automate the categorization, subsequently verifying the accuracy of the mapping through random manual inspections. This family-to-category mapping allowed us to derive the malware classification for each identified family.

Despite this rigorous pipeline, a subset of samples remained unclassified, lacking a definitive family or category. These instances were explicitly assigned an ``unknown'' label at the family and/or category level (\cref{lccd:line:label_unknown}). We deliberately retained these unlabeled samples in the dataset, as the compiled multimodal features and precomputed vector embeddings remain highly valuable for downstream tasks, such as unsupervised clustering, anomaly detection, and exploratory data analysis.

\subsection{Knowledge Enhancement via RAG}\label{subsec:rag}

To optimize LCCD for downstream LLM analysis, we structured the extracted intelligence to support both standard semantic RAG and GraphRAG architectures (\cref{lccd:line:rag_context}). By formatting the dataset to support this dual approach, an analytical framework can retrieve specific definitions via standard RAG while simultaneously using GraphRAG to traverse the complex execution pathways inherent to malware ecosystems.

To populate the RAG context for both implementations with authoritative data, we integrated four specialized cybersecurity databases into our retrieval pipeline:

\begin{itemize}
\item \textbf{MITRE ATT\&CK:} Maps malicious behaviors to standardized Tactics, Techniques, and Procedures (TTPs).
\item \textbf{CWE Vulnerability:} Details software weaknesses and vulnerability patterns.
\item \textbf{Windows API Behavior:} Documents legitimate Windows API usage patterns, providing a baseline to contrast against their malicious implementations in malware.
\item \textbf{Malware Family Intelligence:} Profiles known malware families, detailing attributes ranging from technical capabilities and evasion techniques to historical campaign data and threat actor attribution.
\end{itemize}

Once the relevant context is retrieved from these databases, it is stored in LCCD for later use during downstream inference, and simultaneously injected into the prompt engineering stage described in the subsequent section (\cref{subsec:prompt_engineering}). This dual-purpose pipeline provides the LLM with the verified domain telemetry necessary to accurately execute its analytical tasks.

\subsection{Prompt Engineering and Augmentation}\label{subsec:prompt_engineering}

%%%%%%%%%%%%%%%%%%%%%%%%%%%%%%%%%%%%%%%%%%%%%%%%%%%%%%%%%%%%%%%
% FIGURE: PROMPT ENGINEERING PIPELINE
%%%%%%%%%%%%%%%%%%%%%%%%%%%%%%%%%%%%%%%%%%%%%%%%%%%%%%%%%%%%%%%
\begin{figure*}[pos=htbp]
\centering
\includegraphics[width=0.85\textwidth]{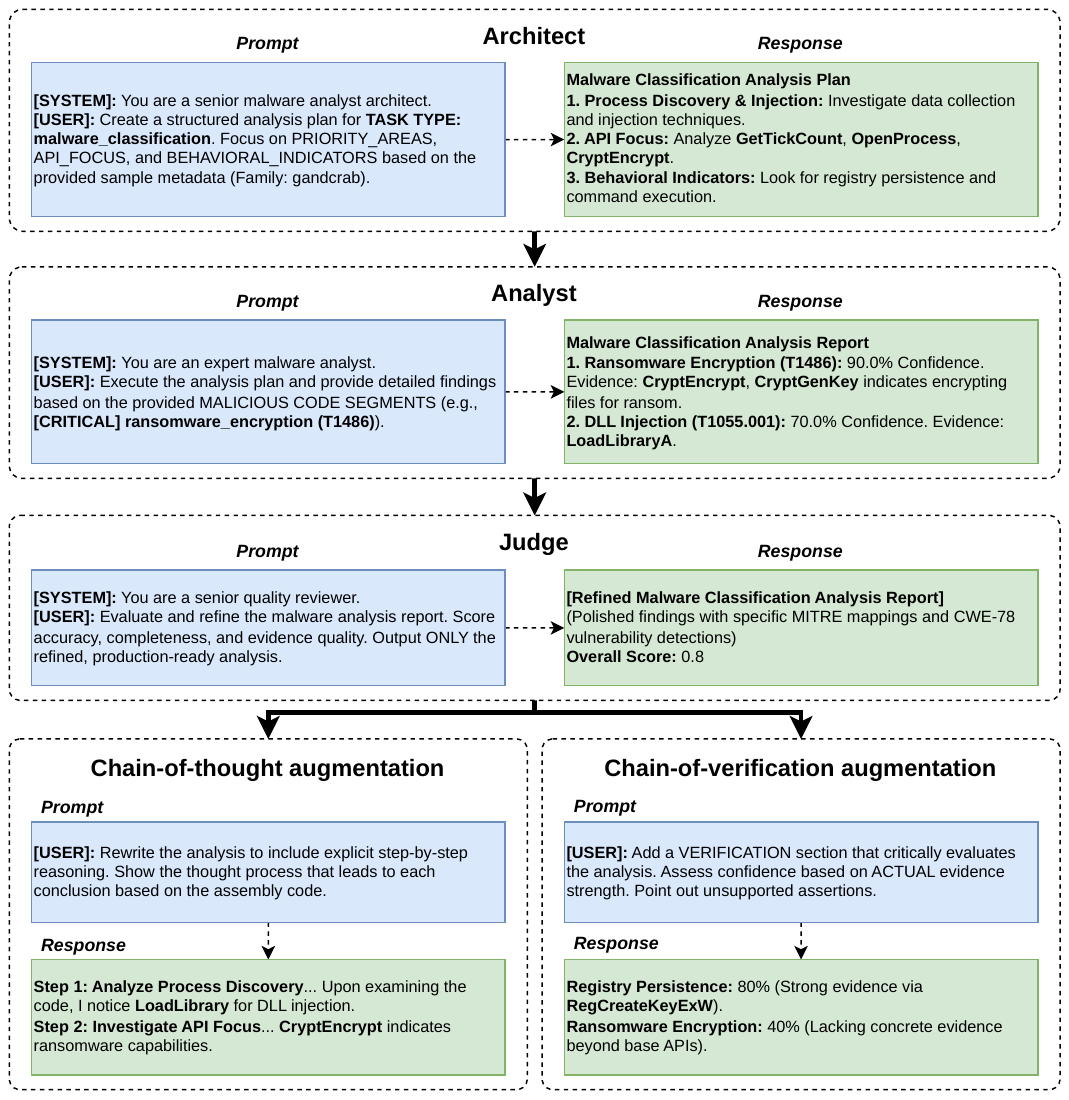}
\caption{Multi-stage prompt engineering pipeline for malware classification. The workflow progresses sequentially through Architect (planning), Analyst (execution), and Judge (evaluation and refinement) roles to generate an initial, high-quality analysis. The refined output is subsequently processed through parallel augmentation modules---Chain-of-Thought and Chain-of-Verification---to enhance reasoning transparency and factual grounding. Note that the prompt and response texts depicted in each node have been significantly condensed for illustrative purposes; the full operational strings contain extensive system context, raw code segments, and strict formatting constraints.}
\label{fig:prompt_pipeline}
\end{figure*}
%==============================================================

To ensure the generation of deep, accurate training data, we implemented a three-stage reasoning and refinement prompt engineering pipeline. This architecture decouples planning from execution and evaluation:
\begin{itemize}
\item\textbf{The Architect:} Processes raw sample data to generate an analysis protocol. It identifies key technical features for extraction, flags potential edge cases for investigation, and outlines the logical steps required to derive a valid conclusion (\cref{lccd:line:prompt_architect}).
\item\textbf{The Analyst:} Executes the Architect’s protocol to produce a comprehensive technical report, mapping the raw data to the specified output format (\cref{lccd:line:prompt_analyst}).
\item\textbf{The Judge:} Performs a comparative audit of the Analyst’s report against the initial plan and the ground truth. It provides qualitative critique and generates the final refined report for inclusion in the training set (\cref{lccd:line:prompt_judge}).
\end{itemize}

Additionally, we crafted diverse system and user prompts for each stage to prevent a model instruction-tuned on this dataset from overfitting to specific syntactic structures. Without this diversity, models risk learning to follow rigid prompt templates, failing to generalize to varied human instructions. These prompts were tailored across 12 fundamental analyst task types, including malware family attribution, behavior and intent analysis, vulnerability and risk assessment, and malware classification.

To further enrich the dataset and instill deliberate reasoning capabilities within a subsequently fine-tuned model, we augmented a subset of the generated training data using CoT~\cite{wei_chain--thought_2022} and CoVe~\cite{dhuliawala_chain--verification_2024} techniques. By applying these methodologies to the finalized outputs of the Judge, we ensured the resulting instruction pairs guide the model to either explicitly decompose the analytical tasks into a series of steps or rigorously verify its own findings before producing a final classification. This complete prompt engineering workflow, including an illustrative example of the generated prompts and responses, is depicted in~\cref{fig:prompt_pipeline}.

Finally, we assigned a difficulty score to all generated prompts based on the inherent characteristics of the analyzed samples. To categorize prompts into \emph{Beginner}, \emph{Intermediate}, or \emph{Expert} tiers, we evaluated three sample-specific metrics: C code length, total function count, and the number of implemented techniques. These metrics directly correlate with analytical complexity: extended code length expands the search space for relevant segments; a high function count obscures control flow tracking, increasing the likelihood of losing execution context; and a dense concentration of techniques complicates the isolation of specific malicious behaviors.

\subsection{Quality Assurance and Validation}\label{subsec:qa_validation}

We integrated an automated quality assurance pipeline to ensure the integrity, reliability, and uniformity of LCCD across five critical dimensions (\cref{lccd:line:validate}):
\begin{itemize}
\item\textbf{Format:} Ensures strict structural compliance, verifying that all generated outputs adhere to expected schemas and predefined templates.
\item\textbf{Content:} Validates completeness and length requirements, ensuring no generated reports are truncated or missing mandatory analytical sections.
\item\textbf{Label:} Audits metadata and classification consistency, confirming that the assigned malware families and categories precisely align with the upstream attribution pipeline. 
\item\textbf{Balance:} Monitors macro-level distribution and diversity metrics across the dataset to flag unintended systemic biases introduced during data collection.
\item\textbf{Quality:} Assesses response quality and formatting indicators, ensuring the LLM-generated text remains technically accurate and logically sound.
\end{itemize}

A unique challenge in generating malware-centric datasets is the frequent triggering of LLM safety guardrails. Because the analytical tasks require parsing malicious code and exploring exploit techniques, models will occasionally refuse to generate a response. During the prompt execution phase, we implemented strict validation checks to detect and reject empty outputs or standard refusal strings (e.g., ``I cannot assist with this request''). We automatically discarded any generation attempt that resulted in a safety refusal and flagged the sample for iterative reprompting, preventing the inclusion of non-informative data.

Furthermore, processing large-scale multimodal data across distributed pipelines inevitably introduces the risk of data loss due to concurrency collisions, API rate limits, or network timeouts. To mitigate this, we implemented robust backfill mechanisms that retroactively query the necessary APIs, regenerate the corrupt representations, or retrigger the LLM generation sequence. This process ensures the finalized dataset is completely exhaustive and strictly consistent.

\subsection{Dataset Statistics}\label{subsec:dataset_statistics}

\subsubsection{Dataset Composition and Extraction Yield}\label{subsubsec:composition_yield}

To evaluate the efficacy of our multimodal extraction pipeline, we analyzed the generation yield across all 34,692 samples in LCCD. As detailed in~\cref{tab:extraction_yield}, the dataset exhibits a highly successful feature extraction rate, reflecting the robustness of our automated backfill and quality assurance mechanisms. The pipeline achieved near-perfect extraction of assembly code via Radare2 (99.95\%), demonstrating the tool's exceptional capacity to disassemble binaries even when decompilers like RetDec fail. While the decompilation of C code via RetDec naturally faced a slightly higher attrition rate (93.84\%) due to the inherent complexities of reversing heavily obfuscated or packed binaries, the intersection of samples successfully processed by both primary tools remained exceptionally high at 93.79\%. Consequently, 92.25\% of the total dataset successfully generated all requisite base embeddings. This comprehensive feature overlap ensures that the vast majority of the dataset contains the complete representations required for advanced downstream LLM tasks.

%%%%%%%%%%%%%%%%%%%%%%%%%%%%%%%%%%%%%%%%%%%%%%%%%%%%%%%%%%%%%%%
% TABLE: DATASET EXTRACTION YIELD AND COMPOSITION
%%%%%%%%%%%%%%%%%%%%%%%%%%%%%%%%%%%%%%%%%%%%%%%%%%%%%%%%%%%%%%%
\begin{table}[pos=htbp]
\scriptsize
    \centering
    \caption{LCCD Extraction Yield and Composition}
    \label{tab:extraction_yield}
    \resizebox{\columnwidth}{!}{%
        \begin{tabular}{@{}llrr@{}}
            \toprule
            \textbf{Category} & \textbf{Metric} & \textbf{Count} & \textbf{Yield Rate (\%)} \\ \midrule
            \multirow{3}{*}{\textbf{Dataset Composition}} 
             & Total Samples & 34,692 & 100.00\% \\
             & Malicious Samples & 33,710 & 97.17\% \\
             & Benign Samples & 982 & 2.83\% \\ \midrule
            \multirow{3}{*}{\textbf{Tooling Success}} 
             & Radare2 Processing & 34,676 & 99.95\% \\
             & RetDec Processing & 32,555 & 93.84\% \\
             & Both Tools Successful & 32,539 & 93.79\% \\ \midrule
            \multirow{5}{*}{\textbf{Extracted Modalities}} 
             & Decompiled C Code & 32,555 & 93.84\% \\
             & Assembly Code & 34,675 & 99.95\% \\
             & FCG \& CFG & 34,271 & 98.79\% \\
             & Complete Base Embeddings & 32,005 & 92.25\% \\ \bottomrule
        \end{tabular}%
    }
\end{table}
%==============================================================

\subsubsection{Instruction-Tuning Data Distribution}\label{subsubsec:instruction_distribution}

To construct a robust instruction-tuning dataset for downstream LLM training, we generated a total of 183,070 sample-prompt pairs. As detailed in~\cref{tab:tuning_distribution}, these pairs are distributed across 12 distinct analytical tasks. The distribution intentionally favors foundational cybersecurity operations, with Malware Classification (13.64\%), Code Analysis (11.32\%), and Malware Class Detection (11.24\%) comprising the largest subsets. This weighting ensures the model develops a strong baseline in fundamental binary analysis before tackling more nuanced, open-ended tasks like Detection Guidance or Intent Analysis.

The generation pipelines were similarly balanced to provide both structural rigidity and reasoning depth. Approximately half of the dataset (53.25\%) was generated using rigorous template-filling to establish strict formatting compliance and baseline factual accuracy. The remaining (46.75\%) leveraged our advanced multi-agent framework. To teach the model step-by-step logic, the final 12.94\% of the dataset was routed through the augmented multi-agent pipeline. We intentionally partitioned this augmented subset into an evenly distributed 1:1:1 ratio across three reasoning configurations: a baseline unaugmented output (4.32\%), Chain-of-Thought (4.31\%), and Chain-of-Verification (4.30\%). This perfectly balanced ablation subset ensures that an instruction-tuned model can learn advanced, multi-step reasoning pathways without artificially overfitting to a single verification methodology.

%%%%%%%%%%%%%%%%%%%%%%%%%%%%%%%%%%%%%%%%%%%%%%%%%%%%%%%%%%%%%%%
% TABLE: DISTRIBUTION OF THE TRAINING DATA
%%%%%%%%%%%%%%%%%%%%%%%%%%%%%%%%%%%%%%%%%%%%%%%%%%%%%%%%%%%%%%%
\begin{table}[pos=htbp]
\scriptsize
    \centering
    \caption{Distribution of the Samples in LCCD}
    \label{tab:tuning_distribution}
    \resizebox{\columnwidth}{!}{%
        \begin{tabular}{@{}lrr@{}}
            \toprule
            \textbf{Category} & \textbf{Total Samples} & \textbf{Percentage (\%)} \\ \midrule
            \textbf{Task Type} & & \\
            \hspace{3mm} Malware Classification & 24,978 & 13.64 \\
            \hspace{3mm} Code Analysis & 20,728 & 11.32 \\
            \hspace{3mm} Malware Class Detection & 20,583 & 11.24 \\
            \hspace{3mm} Malware Family Detection & 14,843 & 8.11 \\
            \hspace{3mm} Risk Assessment & 14,614 & 7.98 \\
            \hspace{3mm} Vulnerability Detection & 14,600 & 7.98 \\
            \hspace{3mm} API Behavior & 14,562 & 7.95 \\
            \hspace{3mm} Threat Identification & 13,854 & 7.57 \\
            \hspace{3mm} Detection Guidance & 11,269 & 6.16 \\
            \hspace{3mm} Family Attribution & 11,266 & 6.15 \\
            \hspace{3mm} Intent Analysis & 11,138 & 6.08 \\
            \hspace{3mm} Technique Explanation & 10,635 & 5.81 \\ \midrule
            \textbf{Generation Pipeline} & & \\
            \hspace{3mm} Template Filling & 97,488 & 53.25 \\
            \hspace{3mm} Architect-Analyst-Judge & 61,897 & 33.81 \\
            \hspace{3mm} Architect-Analyst-Judge (Augmented) & 23,685 & 12.94 \\ \midrule
            \textbf{Reasoning Augmentation} & & \\
            \hspace{3mm} Base Prompting (No Augmentation) & 167,301 & 91.38 \\
            \hspace{3mm} Chain-of-Thought (CoT) & 7,894 & 4.31 \\
            \hspace{3mm} Chain-of-Verification (CoVe) & 7,875 & 4.30 \\ \bottomrule
        \end{tabular}%
    }
\end{table}
%==============================================================

\subsubsection{Class Distribution and Structural Balance} \label{subsubsec:class_distribution}

%%%%%%%%%%%%%%%%%%%%%%%%%%%%%%%%%%%%%%%%%%%%%%%%%%%%%%%%%%%%%%%
% TABLE: TRAINING FRAMEWORK
%%%%%%%%%%%%%%%%%%%%%%%%%%%%%%%%%%%%%%%%%%%%%%%%%%%%%%%%%%%%%%%
\begin{figure*}[pos=htbp]
    \centering
    \includegraphics[width=\linewidth]{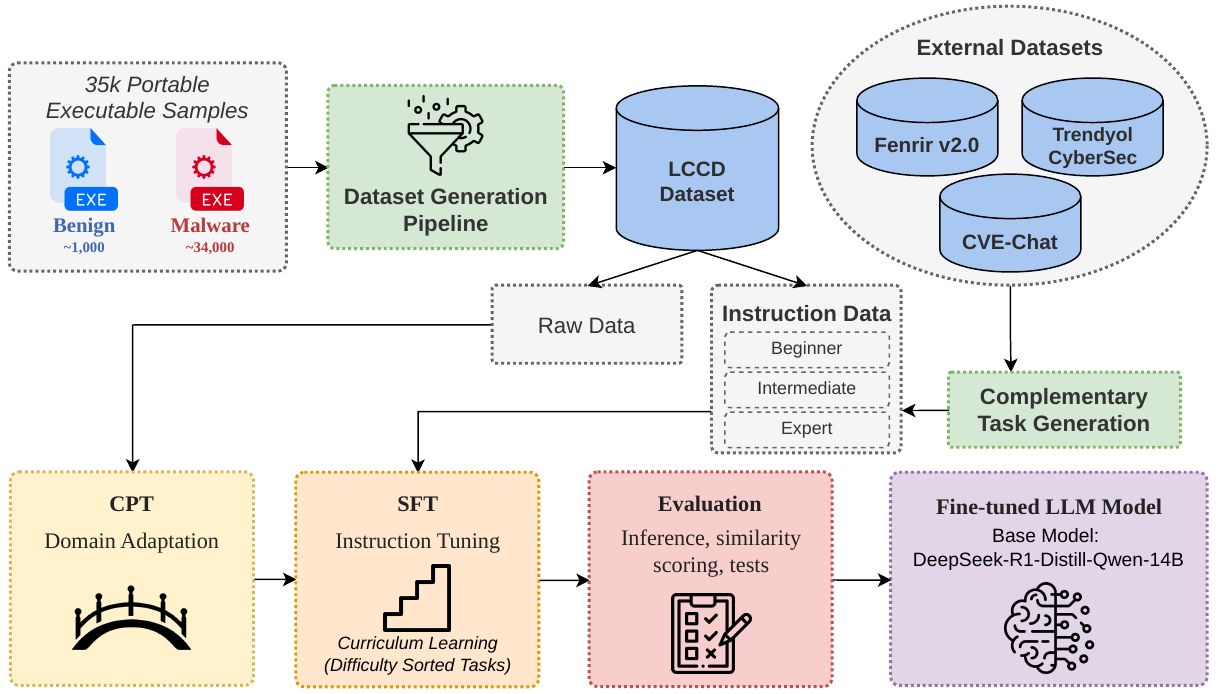}
    \caption{Overview of the proposed training methodology framework. The pipeline illustrates the data flow from raw Portable Executable (PE) samples through the Dataset Generation Pipeline to  the LCCD. The extracted intelligence is partitioned into unstructured Raw Data for Continued Pre-Training (CPT) and stratified Instruction Data. The Instruction Data is augmented with external datasets and fed into the Supervised Fine-Tuning (SFT) phase utilizing a Curriculum Learning schedule to incrementally adapt the foundational DeepSeek-R1-Distill-Qwen-14B model.}
    \label{fig:framework_diagram}
\end{figure*}
%==============================================================

LCCD has 624 different malware families represented spread across 21 malware categories. There exists a class imbalance driven by the real-world distribution in which in-the-wild malware is not balanced. Malware campaigns can dominate the threat landscape for long periods of time leading to a bigger sample size of that respective family. We intentionally preserved the inherent class imbalance to ensure an authentic representation of the malware ecosystem. Nonetheless, we still evaluated the structural balance by calculating the Shannon Entropy ($H$) for each category. This value lets us measure the uncertainty and diversity in the categories with respect to the malware families and the samples belonging to each of them. The Shannon Entropy is calculated as:

\noindent
\begin{minipage}{\columnwidth}
\begin{equation}
    \label{eq:shannon_entropy}
    H = -\sum_{i=1}^{S} p_i \log_2 p_i
\end{equation}
where $S$ is the total number of distinct malware families and $p_i$ the proportion of samples in the $i$-th family
\end{minipage}

% \begin{equation}
%     \label{eq:shannon_entropy}
%     H = -\sum_{i=1}^{S} p_i \log_2 p_i
% \end{equation}

% \noindent where:
% \begin{center}
%     \begin{tabular}{ r @{ : \quad} l }
%         $S$   & total number of distinct malware families \\
%         $p_i$ & proportion of samples in the $i$-th family \\
%     \end{tabular}
% \end{center}

A higher $H$ value indicates greater diversity among the families within that category, while a category dominated by a single family will have a score very close to zero. However, the Shannon entropy is sensitive to the total number of families ($S$), and in LCCD, the ratio of families per category varies drastically. Because the maximum possible entropy ($H_{max}$) increases logarithmically with $S$, directly comparing the Shannon Entropy between categories would not give a fair view of relative balance.

To establish a standardized metric for comparison among the categories, we utilize Pielou’s Evenness ($J'$). This metric normalizes the Shannon Entropy $H$ by dividing it by the maximum possible entropy for that specific category’s family count $S$:
\begin{equation}
    \label{eq:pielou_evenness}
    J' = \frac{H}{\log_2 S}
\end{equation}

A Pielou’s Evenness score of 1 indicates a perfectly uniform distribution of samples across all families in the category, whereas a score approaching 0 indicates an imbalance dominated by a single family. \cref{tab:malware_category_dist} depicts the malware categories present in the dataset, organized by their number of samples. To depict the dataset distribution, we included the top family belonging to the category alongside the proportion it represents. In the case of categories in which an unknown label dominates, we included the second top family for reference. These last cases happen when we were able to label the samples with a category but not pinpoint the exact malware family to which the sample belongs. The last column shows Pielou’s Evenness values, except for the ``Benign'' category, which is omitted as it contains only a single class.

%%%%%%%%%%%%%%%%%%%%%%%%%%%%%%%%%%%%%%%%%%%%%%%%%%%%%%%%%%%%%%%
% TABLE: CLASS DIVERSITY
%%%%%%%%%%%%%%%%%%%%%%%%%%%%%%%%%%%%%%%%%%%%%%%%%%%%%%%%%%%%%%%
\begin{table*}[pos=t]
\scriptsize
    \centering
    \caption{Intra-Class Diversity and Pielou's Evenness Across LCCD Malware Categories}
    \label{tab:malware_category_dist}
    \begin{tabular*}{\textwidth}{@{\extracolsep{\fill}} l c c l c c @{}}
        \toprule
        \textbf{Category} & \textbf{Total Samples} & \textbf{Total Families ($S$)} & \textbf{Top Family} & \textbf{Top Family \%} & \textbf{Pielou's Evenness ($J'$)} \\
        \midrule
        Rat & 6,026 & 89 & AsyncRAT & 32.8\% & 0.481 \\
        Ransomware & 5,720 & 105 & Gandcrab & 58.5\% & 0.300 \\
        Stealer & 4,546 & 94 & AgentTesla & 22.9\% & 0.631 \\
        Rogueware & 4,401 & 2 & Winwebsec & 100.0\% & 0.003 \\
        \addlinespace
        Loader & 3,291 & 72 & Unknown/Amadey & 54.4\% / 7.4\% & 0.473 \\
        Banker & 2,296 & 21 & Zbot & 91.6\% & 0.161 \\
        Unknown & 1,653 & 18 & Unknown/Boxter & 98.6\% / 0.1\% & 0.039 \\
        Backdoor & 1,633 & 68 & Unknown/Meterpreter & 21.4\% / 18.0\% & 0.619 \\
        \addlinespace
        Adware & 1,555 & 23 & Mediyes & 93.2\% & 0.128 \\
        Botnet & 1,512 & 50 & ZeroAccess & 45.8\% & 0.520 \\
        Benign & 982 & 1 & Benign & 100.0\% & N/A \\
        Trojan & 363 & 38 & Casdet & 11.3\% & 0.855 \\
        \addlinespace
        Miner & 189 & 7 & Coinminer & 61.9\% & 0.576 \\
        Hacktool & 136 & 26 & Cobalt & 30.9\% & 0.759 \\
        Dropper & 99 & 7 & Babadeda & 65.7\% & 0.550 \\
        Worm & 85 & 14 & Mofksys & 69.4\% & 0.497 \\
        \addlinespace
        Keylogger & 69 & 4 & VIP & 79.7\% & 0.519 \\
        Virus & 51 & 8 & Neshta & 78.4\% & 0.442 \\
        Wiper & 47 & 10 & KillMBR & 29.8\% & 0.889 \\
        Spyware & 21 & 5 & Kimsuky & 47.6\% & 0.766 \\
        \addlinespace
        Rootkit & 10 & 6 & Nemesis & 30.0\% & 0.917 \\
        Grayware & 4 & 2 & AdLoad & 75.0\% & 0.811 \\
        Exploit & 3 & 2 & JuicyPotato & 66.7\% & 0.918 \\
        \bottomrule
    \end{tabular*}
\end{table*}
%==============================================================

%~~~~~~~~~~~~~~~~~~~~~~~~~~~~~~~~~~~~~~~~~~~~~~~~~~~~~~~~~~~~~~~~~~~~~~~~~~~~~~~~~~~~~~~~
\section{Proposed Methodology}\label{sec:proposed_methodology}

This section presents the proposed LCC-LLM methodology for code-centric malware attribution and multi-task static malware analysis. As shown in~\cref{fig:framework_diagram}, the framework begins with approximately 35K PE samples, including benign and malicious executables, which are processed through the LCCD generation pipeline. LCCD is organized into two main streams: raw data and instruction data. The raw data stream supports Continued Pre-Training (CPT) for domain adaptation, while the instruction data stream supports Supervised Fine-Tuning (SFT) using curriculum learning with beginner, intermediate, and expert-level tasks.

In addition, external cybersecurity datasets, including Fenrir v2.0, Trendyol CyberSec, and CVE-Chat, are used for complementary task generation to improve task diversity and domain coverage. The resulting instruction corpus is used to adapt DeepSeek-R1-Distill-Qwen-14B~\cite{guo_deepseek-r1_2025} for malware attribution, malicious code segment identification, vulnerability analysis, IoC extraction, MITRE ATT\&CK mapping, and structured report generation. To reduce training overhead, the model is fine-tuned using QLoRA, enabling efficient adaptation while preserving the reasoning and code-understanding capabilities of the base model.

\begin{algorithm}[t]
\caption{Evidence-Grounded LCC-LLM Inference Framework}
\scriptsize
\label{alg:lcc_inference}
\begin{algorithmic}[1]
    \Require User query $Q$, optional malware sample $P$, fine-tuned model $M$
    \Ensure Analyst-ready response $R$ and optional structured report

    \State $T \gets \emptyset$ \Comment{Static-analysis transcript} \label{lcc_inference:line:init_start}
    \State $R \gets \emptyset$ \Comment{Final analyst response}
    \State $Q_n \gets \text{Normalize}(Q)$
    \State $\kappa \gets H(Q_n,P,M)$ \Comment{Cache key} \label{lcc_inference:line:init_end}

    \If{$\text{CacheHit}(\kappa)$}
        \State \Return $\text{CachedResponse}(\kappa)$
    \EndIf

    \If{$P \neq \emptyset$} \label{lcc_inference:line:static_analysis_chain_start}
        \State $\tau \gets \text{DetectFileTypeByMagicBytes}(P)$
        \State $T \gets \text{RunStaticToolChain}(P,\tau)$
    \EndIf \label{lcc_inference:line:static_analysis_chain_end}

    \State $\Gamma \gets \text{HybridRetrieve}(Q_n,T)$ \label{lcc_inference:line:prompt_start}
    \State $Q^{*} \gets \text{FormatPrompt}(Q_n,\Gamma,T)$ \label{lcc_inference:line:prompt_end}

    \State $\sigma \gets \text{MatchSpecialist}(Q^{*})$ \label{lcc_inference:line:response_generation_start}
    \If{$\sigma \neq \emptyset$}
        \State $R \gets \text{RunSpecialistAgent}(\sigma,Q^{*},\Gamma,T)$
    \Else
        \State $R \gets \text{GenerateWithLLM}(M,Q^{*})$
    \EndIf \label{lcc_inference:line:response_generation_end}

    \State $R \gets \text{VerifyAndValidate}(R,\Gamma,T)$ \label{lcc_inference:line:verification}

    \If{$P \neq \emptyset$} \label{lcc_inference:line:report_start}
        \State $R \gets \text{AttachStructuredEvidence}(R,T)$
        \State $\mathcal{P}_{rep} \gets \text{GenerateStructuredReport}(P,R,T)$
    \EndIf \label{lcc_inference:line:report_end}

    \State $\text{CacheStore}(\kappa,R)$
    \State \Return $R$
\end{algorithmic}
\end{algorithm}

\begin{algorithm}[t]
\caption{Hybrid Retrieval and Static Tool Orchestration}
\scriptsize
\label{alg:rag_tool}
\begin{algorithmic}[1]
    \Require Query $Q$, optional sample transcript $T$, retrieval collections $\mathcal{C}$
    \Ensure Grounded context bundle $\Gamma$

    \State $\Gamma \gets \emptyset$ \label{rag_tool:line:init_start}

    \For{each collection $C_i \in \mathcal{C}$} \label{rag_tool:line:init_end}
        \State $S_{\mathrm{BM25}} \gets \text{BM25Search}(Q,C_i)$ \label{rag_tool:line:bm25search}
        \State $S_{\mathrm{dense}} \gets \text{DenseSearch}(\text{Encode}(Q),C_i)$ \label{rag_tool:line:densesearch}
        \State $S_{\mathrm{rrf}} \gets \text{ReciprocalRankFusion}(S_{\mathrm{BM25}},S_{\mathrm{dense}})$ \label{rag_tool:line:combinedrefined_start}
        \State $S_{\mathrm{ranked}} \gets \text{CrossEncoderRerank}(Q,S_{\mathrm{rrf}})$
        \State $\Gamma \gets \Gamma \cup \text{TopK}(S_{\mathrm{ranked}})$ \label{rag_tool:line:combinedrefined_end}
    \EndFor

    \If{$T \neq \emptyset$} \label{rag_tool:line:enrichment_start}
        \State $E_c \gets \text{ExtractCodeEvidence}(T)$
        \Comment{decompiled code, assembly, CFG/FCG}
        \State $E_s \gets \text{ExtractSecurityEvidence}(T)$
        \Comment{suspicious APIs, Capa matches, PE metadata}
        \State $\Gamma \gets \Gamma \cup E_c \cup E_s$
    \EndIf \label{rag_tool:line:enrichment_end}

    \State $\Gamma \gets \text{RemoveDuplicates}(\Gamma)$ \label{rag_tool:line:cleanup_start}
    \State $\Gamma \gets \text{FilterLowConfidence}(\Gamma)$ 
    \State \Return $\Gamma$ \label{rag_tool:line:cleanup_end}
\end{algorithmic}
\end{algorithm}

\cref{alg:lcc_inference} summarizes the overall inference procedure of the proposed LCC-LLM framework. The process begins by initializing the response state and static analysis transcript, followed by query normalization and cache-key generation (\crefrange{lcc_inference:line:init_start}{lcc_inference:line:init_end}). This step avoids redundant computation when the same query, model, and malware sample have already been analyzed. If a malware sample is provided, the framework identifies the file type through magic-byte inspection and executes the corresponding static-analysis chain (\crefrange{lcc_inference:line:static_analysis_chain_start}{lcc_inference:line:static_analysis_chain_end}). The resulting transcript contains the reverse-engineering evidence required for grounded analysis, including PE metadata, decompiled code, assembly, CFG/FCG artifacts, suspicious APIs, and tool-generated findings. The retrieved knowledge and tool-derived evidence are then integrated into a grounded context bundle $\Gamma$ and used to construct an evidence-enriched prompt $Q^{*}$ (\crefrange{lcc_inference:line:prompt_start}{lcc_inference:line:prompt_end}). Based on the query intent, the framework either invokes a specialist agent or the fine-tuned LCC-LLM model for response generation (\crefrange{lcc_inference:line:response_generation_start}{lcc_inference:line:response_generation_end}). The generated response is not returned directly; instead, it is passed through verification and validation to reduce unsupported claims and improve factual reliability (\cref{lcc_inference:line:verification}). Finally, when a malware sample is available, structured evidence is attached to the response and an analyst-oriented report can be generated (\crefrange{lcc_inference:line:report_start}{lcc_inference:line:report_end}). This design ensures that the final output is not only generated by an LLM, but also grounded in retrieved cybersecurity knowledge and deterministic reverse-engineering evidence.

Next, \cref{alg:rag_tool} describes the retrieval and evidence-construction mechanism used to support grounded malware reasoning. The algorithm first initializes an empty context bundle $\Gamma$ and iterates over the available cybersecurity knowledge collections (\crefrange{rag_tool:line:init_start}{rag_tool:line:init_end}). For each collection, lexical retrieval based on BM25 is applied to capture exact security identifiers, such as CVE IDs, CWE entries, MITRE ATT\&CK technique, hashes, and API names (\cref{rag_tool:line:bm25search}). In parallel, dense retrieval is used to identify semantically related malware behaviors, threat descriptions, and contextual evidence that may not share exact keywords with the query (\cref{rag_tool:line:densesearch}). The two retrieval outputs are then combined using reciprocal-rank fusion and refined using cross-encoder reranking to select the most relevant evidence (\crefrange{rag_tool:line:combinedrefined_start}{rag_tool:line:combinedrefined_end}). When a static-analysis transcript is available, the retrieval context is further enriched with code-level and security-level evidence extracted directly from the analyzed sample (\crefrange{rag_tool:line:enrichment_start}{rag_tool:line:enrichment_end}). This includes decompiled code, assembly, CFG/FCG artifacts, suspicious API evidence, Capa matches, and PE metadata. Finally, duplicate and low-confidence contexts are removed before returning the final context bundle (\crefrange{rag_tool:line:cleanup_start}{rag_tool:line:cleanup_end}). This step is important because it enables the framework to combine external cybersecurity knowledge with sample-specific reverse-engineering evidence, thereby reducing reliance on unsupported LLM generation.

\begin{algorithm}[t]
\scriptsize
\caption{Verification-Guided Output Validation}
\label{alg:verification_quality}
\begin{algorithmic}[1]
    \Require Generated response $R$, grounded context $\Gamma$, tool transcript $T$
    \Ensure Verified and quality-gated response $R^{*}$

    \State $I \gets \text{ExtractIndicators}(R)$ 
    \Comment{MITRE, CVE, CWE, CAPEC, CVSS, IP, hash, URL, email, port} \label{verification_quality:line:extract}

    \For{each indicator $i \in I$} \label{verification_quality:line:annotate_start}
        \State $v_i \gets \text{ValidateIndicator}(i)$
        \State $R \gets \text{AttachProvenanceLabel}(R,i,v_i)$
        \Comment{verified, valid-unverified, or invalid}
    \EndFor \label{verification_quality:line:annotate_end}

    \State $d_1 \gets \text{InformationDensity}(R)$ \label{verification_quality:line:evaluation_start}
    \State $d_2 \gets \text{StructuralCompleteness}(R)$
    \State $d_3 \gets \text{RepetitionPenalty}(R)$
    \State $d_4 \gets \text{LengthSanity}(R)$
    \State $d_5 \gets \text{EvidenceAlignment}(R,\Gamma,T)$
    \State $\sigma \gets \text{WeightedQualityScore}(d_1,d_2,d_3,d_4,d_5)$ \label{verification_quality:line:evaluation_end}

    \If{$\sigma \geq \tau_{\mathrm{accept}}$} \label{verification_quality:line:qualityscore_start}
        \State $R^{*} \gets R$
    \ElsIf{$\sigma \geq \tau_{\mathrm{retry}}$}
        \State $R^{*} \gets \text{RegenerateWithFeedback}(R,\Gamma,T)$
    \Else
        \State $R^{*} \gets \text{GenerateFromValidatedTemplate}(\Gamma,T)$
    \EndIf \label{verification_quality:line:qualityscore_end}

    \State $R^{*} \gets \text{AttachVerifiedEvidence}(R^{*},\Gamma,T)$
    \Comment{IoCs, MITRE mappings, suspicious APIs, code-level evidence} \label{verification_quality:line:enrich_start}
    \State \Return $R^{*}$ \label{verification_quality:line:enrich_end}
\end{algorithmic}
\end{algorithm}

\cref{alg:verification_quality} presents the verification-guided validation stage applied after response generation. The algorithm first extracts security-relevant indicators from the generated response, including MITRE ATT\&CK techniques, CVEs, CWEs, CAPEC entries, CVSS vectors, IP addresses, hashes, URLs, email addresses, and ports (\cref{verification_quality:line:extract}). Each extracted indicator is then validated and annotated with a provenance label indicating whether it is verified, valid-but-unverified, or invalid (\crefrange{verification_quality:line:annotate_start}{verification_quality:line:annotate_end}). This mechanism is designed to reduce hallucinated IoCs, incorrect mappings, and unsupported attribution claims. In addition to indicator validation, the response is evaluated using five quality dimensions: information density, structural completeness, repetition penalty, length sanity, and evidence alignment (\crefrange{verification_quality:line:evaluation_start}{verification_quality:line:evaluation_end}). These dimensions are combined into a weighted quality score $\sigma$, which determines whether the response should be accepted, regenerated with feedback, or reconstructed from a validated template (\crefrange{verification_quality:line:qualityscore_start}{verification_quality:line:qualityscore_end}). The final response is then enriched with verified IoCs, MITRE mappings, suspicious APIs, and code-level evidence before being returned to the analyst (\crefrange{verification_quality:line:enrich_start}{verification_quality:line:enrich_end}). This validation process strengthens the reliability of LCC-LLM by ensuring that the final output is traceable, evidence-aligned, and suitable for analyst-facing malware attribution workflows.

\subsection{Infrastructure}\label{subsec:infrastructure}
All training and evaluation experiments were conducted on the IBEX High-Performance Computing (HPC) cluster at King Abdullah University of Science and Technology (KAUST). The cluster provides access to heterogeneous GPU resources; however, as a shared academic platform, experimental design must balance computational capacity, job-queue availability, and allocation constraints.

For the main training configuration, we used a single SLURM compute-node equipped with four NVIDIA A100 GPUs, each with 80\,GB of memory, together with 300\,GB of system RAM and 32~CPU cores, as summarized in~\cref{tab:slurm-allocation}. This configuration provides 320\,GB of aggregate GPU memory and supports efficient fine-tuning of the selected 14-billion-parameter backbone model. To reduce the computational and memory overhead, we adopted 4-bit QLoRA, which freezes the quantized base model and updates only a small number of trainable adapter parameters. This configuration enables memory-efficient domain adaptation while preserving the reasoning and code-understanding capabilities of the base model. The use of QLoRA substantially reduces the hardware requirements compared with full-parameter fine-tuning and improves the practicality of reproducing the proposed training pipeline on moderate multi-GPU academic infrastructure. Nevertheless, larger batch sizes, faster convergence, and shorter wall-clock training times benefit from high-memory GPUs such as the A100.

%%%%%%%%%%%%%%%%%%%%%%%%%%%%%%%%%%%%%%%%%%%%%%%%%%%%%%%%%%%%%%%
% TABLE: SLURM RESOURCE ALLOCATION
%%%%%%%%%%%%%%%%%%%%%%%%%%%%%%%%%%%%%%%%%%%%%%%%%%%%%%%%%%%%%%%
\begin{table}[pos=htbp]
    \centering
    \scriptsize
    \caption{SLURM resource allocation for the $1\times4$ A100 configuration.}
    \vspace{1em}
    \begin{tabular}{ll}
        \toprule
        \textbf{Resource} & \textbf{Value} \\
        \midrule
        Nodes            & 1 \\
        GPUs per node    & $4 \times$ NVIDIA A100 (80\,GB) \\
        CPUs per task    & 32 \\
        Memory per node  & 300\,GB \\
        \bottomrule
    \end{tabular}
    \label{tab:slurm-allocation}
\end{table}
%==============================================================

\subsection{Implementation Details}\label{subsec:framework_implementation_details}

%%%%%%%%%%%%%%%%%%%%%%%%%%%%%%%%%%%%%%%%%%%%%%%%%%%%%%%%%%%%%%%
% FIGURE: TASK TYPE FOR PROMPT GENERATION DURING FINE TUNING
%%%%%%%%%%%%%%%%%%%%%%%%%%%%%%%%%%%%%%%%%%%%%%%%%%%%%%%%%%%%%%%
\begin{figure}[pos=htbp]
    \centering
    \includegraphics[width=\columnwidth]{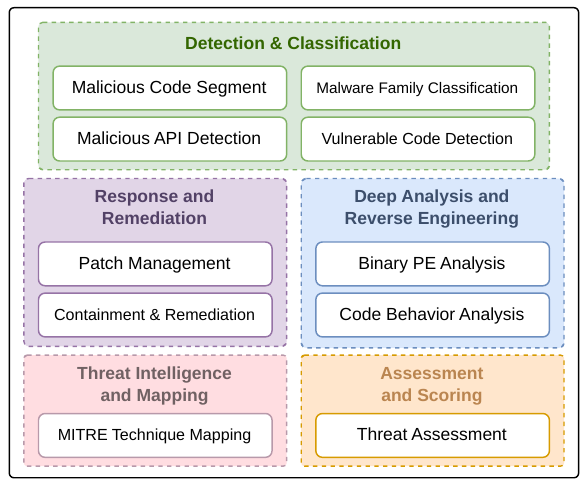}
    \caption{The 10 core task types implemented in the task generator. The tasks are grouped into five distinct categories covering the threat analysis lifecycle, from initial detection through deep analysis and remediation.}
    \label{fig:task_type_generation}
\end{figure}
%==============================================================

%%%%%%%%%%%%%%%%%%%%%%%%%%%%%%%%%%%%%%%%%%%%%%%%%%%%%%%%%%%%%%%
% TABLE: CPT HYPERPARAMETERS
%%%%%%%%%%%%%%%%%%%%%%%%%%%%%%%%%%%%%%%%%%%%%%%%%%%%%%%%%%%%%%%
\begin{table}[pos=htbp]
    \centering
    \scriptsize
    \caption{CPT Training Configuration}
    \vspace{0.5em}
    \begin{tabular}{ll}
        \toprule
        \textbf{Parameter} & \textbf{Value} \\
        \midrule
        Epochs & 3 (early stopped at $\sim$1.2) \\
        Learning Rate & 1e-5 (cosine decay) \\
        Warmup Ratio & 5\% \\
        Effective Batch Size & 32 (4 $\times$ 2 $\times$ 4 GPUs) \\
        Max Grad Norm & 1.0 \\
        Weight Decay & 0.01 \\
        Optimizer & AdamW (fused) \\
        Gradient Checkpointing & Enabled \\
        DeepSpeed & ZeRO-2 \\
        Total Steps & 4,200 \\
        Eval Steps & 100 \\
        Save Steps & 200 \\
        Save Total Limit & 3 \\
        Final Train Loss & 0.9224 \\
        Final Eval Loss & 1.1770 \\
        Final Train Accuracy & 0.8027 \\
        \bottomrule
    \end{tabular}
    \label{tab:cpt_training_configuration}
\end{table}
%==============================================================

%%%%%%%%%%%%%%%%%%%%%%%%%%%%%%%%%%%%%%%%%%%%%%%%%%%%%%%%%%%%%%%
% TABLE: SFT HYPERPARAMETERS
%%%%%%%%%%%%%%%%%%%%%%%%%%%%%%%%%%%%%%%%%%%%%%%%%%%%%%%%%%%%%%%
\begin{table}[pos=htbp]
    \centering
    \scriptsize
    \caption{SFT Training Configuration}
    \vspace{0.5em}
    \begin{tabular}{ll}
        \toprule
        \textbf{Parameter} & \textbf{Value} \\
        \midrule
        Epochs & 2 \\
        Learning Rate & 5e-5 (peak, cosine decay) \\
        Warmup & Warm-up to peak LR \\
        Effective Batch Size & 16 (2 $\times$ 4 $\times$ 2 GPUs) \\
        Max Grad Norm & 0.5 \\
        Weight Decay & 0.01 \\
        Optimizer & AdamW (fused) \\
        NEFTune alpha & 10.0 \\
        Curriculum Learning & Enabled (difficulty-sorted) \\
        Sequence Packing & Enabled (max 4096 tokens) \\
        Total Steps & 1,550 \\
        Eval Steps & 100 \\
        Final Train Loss & 0.1471 \\
        Final Eval Loss & 0.1541 \\
        Final Train Accuracy & 0.9539 \\
        Final Eval Accuracy & 0.9537 \\
        \bottomrule
    \end{tabular}
    \label{tab:sft_training_configuration}
\end{table}
%==============================================================

\textbf{Pipeline Objectives:} The combination of CPT and SFT in our training pipeline systematically adapts the foundational model to the malware analysis domain before aligning it to execute specific analyst workflows.

\textbf{Data Partitioning and Export:}\label{paragraph:data_importing} From LCCD, we exported two distinct data partitions as Parquet files to leverage their native compression and the fast query times enabled by columnar projection. These partitions served distinct purposes across the successive stages of our model fine-tuning. For the CPT stage (\cref{paragraph:cpt_implementation}) and data preprocessing (\cref{paragraph:data_preprocessing}), we utilized data from approximately 4,000 raw samples, encompassing the code variants (\cref{subsec:multimodal_representations}), metadata, and static analysis artifacts (\cref{subsec:threat_intelligence}). For the SFT stage, we instead sampled 8,000 System/User/Assistant instruction triplets. Although LCCD contains a significantly larger volume of generated data, we restricted the training sample size to accommodate computational constraints. While reducing the number of training samples may incrementally impact peak accuracy, this sampling strategy provides an optimal trade-off between maintaining robust model performance and ensuring highly efficient training cycles.

\textbf{Data Preprocessing:}\label{paragraph:data_preprocessing} Using the raw sample data, we dynamically generated additional instruction-response pairs (based on the tasks outlined in~\cref{fig:task_type_generation}) to augment the data statically generated during the initial pipeline (\cref{sec:methodology_dataset}). We mixed these dynamic and static sets of training data to achieve three primary objectives. First, this mixture prevents the model from over-fitting to the specific syntactic templates of the static data. Second, it increases reasoning path diversity, forcing the model to learn alternative methods for reaching the same analytical conclusion. Finally, this mixture strikes a balance between high-quality static ``anchor'' pairs, which ground the model against hallucinations, and dynamic ``explorer'' pairs that introduce the variance necessary for robust generalization. The merged training data was then passed through a deduplication and normalization pipeline. This ensured schema consistency and filtered out any samples failing our quality criteria prior to splitting the data into the final training, evaluation, and test sets.

%%%%%%%%%%%%%%%%%%%%%%%%%%%%%%%%%%%%%%%%%%%%%%%%%%%%%%%%%%%%%%%
% FIGURE: CPT DYNAMICS
%%%%%%%%%%%%%%%%%%%%%%%%%%%%%%%%%%%%%%%%%%%%%%%%%%%%%%%%%%%%%%%
\begin{figure*}[pos=htbp]
    \centering
    \includegraphics[width=0.85\textwidth]{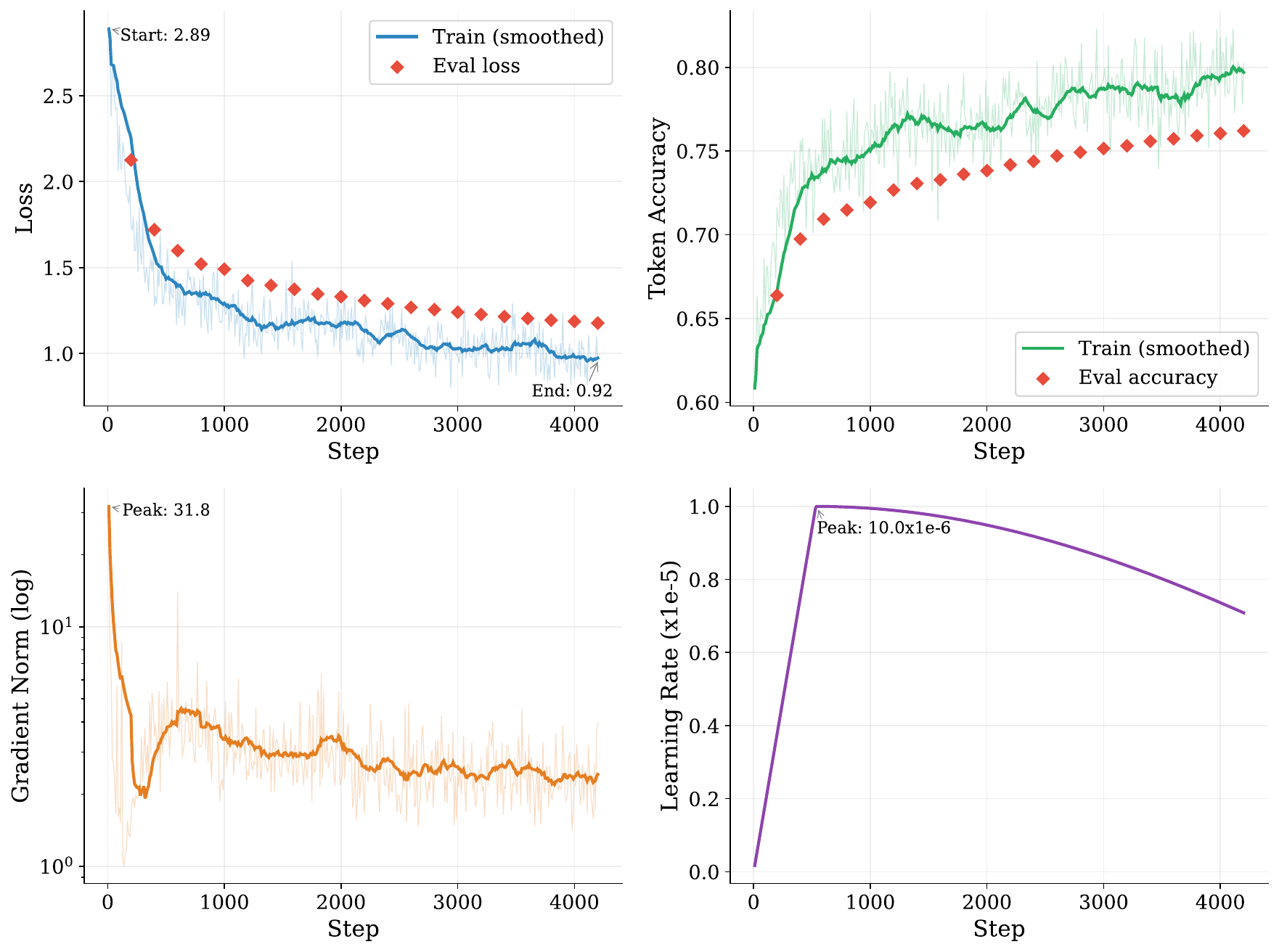} 
    \caption{Continued Pre-Training (CPT) dynamics over 4,200 optimization steps. \textbf{(A)} Training and evaluation loss curves, demonstrating a steady reduction from an initial 2.89 to a final 0.92. \textbf{(B)} Mean token accuracy, which increased from 0.609 to 0.803, indicating successful adaptation to the vocabulary and syntax of the decompiled code. \textbf{(C)} Gradient norm progression. An initial gradient spike (31.8) stabilized rapidly within the first 100 steps, confirming the efficacy of the learning rate warm-up phase. \textbf{(D)} The learning rate schedule, featuring a linear warm-up peaking at $1.0 \times 10^{-5}$ followed by a cosine decay.}
    \label{fig:cpt_dynamics}
\end{figure*}
%==============================================================

\textbf{Continued Pre-Training:} \label{paragraph:cpt_implementation} CPT performs the initial domain adaptation by training the model directly on raw executable representations (detailed hyperparameter configurations are provided in~\cref{tab:cpt_training_configuration}). This unsupervised phase teaches the model the unique vocabulary, syntax patterns, and structural conventions of reverse-engineered code. It is imperative to perform this adaptation prior to task-specific instruction tuning to ensure the binary artifacts included in the prompts are accurately parsed by the model.

\textbf{Supervised Fine-Tuning:} \label{paragraph:sft_implementation} Building upon the CPT foundation, SFT aligns the model to follow complex malware analysis directives. The model is trained on our curated dataset using ChatML-formatted prompts and tasks generated from external datasets Fenrir v2.0~\cite{kiraz_fenrir_2025}, Trendyol Cybersecurity Defense Instruction-Tuning~\cite{trendyol_security_team_trendyol_2025}, and CVE Chat-Style Multi-Turn Cybersecurity~\cite{ansulev_cve_2026} using the configuration outlined in~\cref{tab:sft_training_configuration}. With this training, the model learns to generate structured intelligence reports, map identified behaviors to specific techniques, and analyze decompiled code to identify vulnerabilities and malicious segments.

\textbf{Evaluation:} After the CPT and SFT stages, an evaluation stage follows, starting with a subset of 30 samples. These samples are assessed to determine if they hit a mean semantic similarity score of at least 0.70 with respect to reference responses, serving as an initial checkpoint designed to measure the baseline quality of the model. The evaluation framework then executes an evaluation across 43 distinct task types using a Hugging Face pipeline. Performance is measured using a dual-metric approach: every response is scored for overall semantic similarity using the all-MiniLM-L6-v2 SentenceTransformer, while task-specific functions extract additional metrics. To ensure the model is practical for the analysis of real-world scenarios, the framework subjects the model to advanced testing across five complex scenarios. Each scenario presents a multi-step analysis challenge that requires the model to synthesize domain knowledge across multiple task types. The proposed scenarios are: Code Injection Analysis, Trojanized PE Detection, Multi-Vuln SAST, Family Classification, and Ransomware Containment.

%~~~~~~~~~~~~~~~~~~~~~~~~~~~~~~~~~~~~~~~~~~~~~~~~~~~~~~~~~~~~~~~~~~~~~~~~~~~~~~~~~~~~~~~~
\section{Experimental Results and Discussion}\label{sec:experimental_results}

\textbf{Continued Pre-Training:} \label{paragraph:cpt_results} The CPT phase successfully adapted the base model to the specialized domain of malware analysis, as illustrated by the training dynamics in~\cref{fig:cpt_dynamics}. Over the course of approximately 4,200 optimization steps, the model exhibited steady and stable convergence. The training and evaluation loss decreased significantly from an initial value of 2.89 to 0.92, indicating that the model effectively internalized the underlying structure of the dataset with the raw samples. This domain adaptation is further corroborated by the token accuracy, which climbed from a baseline of roughly 0.61 to over 0.80. This steady increase demonstrates the model's growing proficiency in parsing and predicting the unique vocabulary and syntactic patterns of the diverse code representations of a malware sample. Furthermore, despite the inherent complexity of unstructured binary artifacts, the training regime remained highly stable; an initial gradient norm spike of 31.8 was rapidly mitigated within the first 100 steps, proving the efficacy of the linear learning-rate warm-up phase.

%%%%%%%%%%%%%%%%%%%%%%%%%%%%%%%%%%%%%%%%%%%%%%%%%%%%%%%%%%%%%%%
% FIGURE: SFT DYNAMICS
%%%%%%%%%%%%%%%%%%%%%%%%%%%%%%%%%%%%%%%%%%%%%%%%%%%%%%%%%%%%%%%
\begin{figure*}[pos=htbp]
    \centering
    \includegraphics[width=0.80\textwidth]{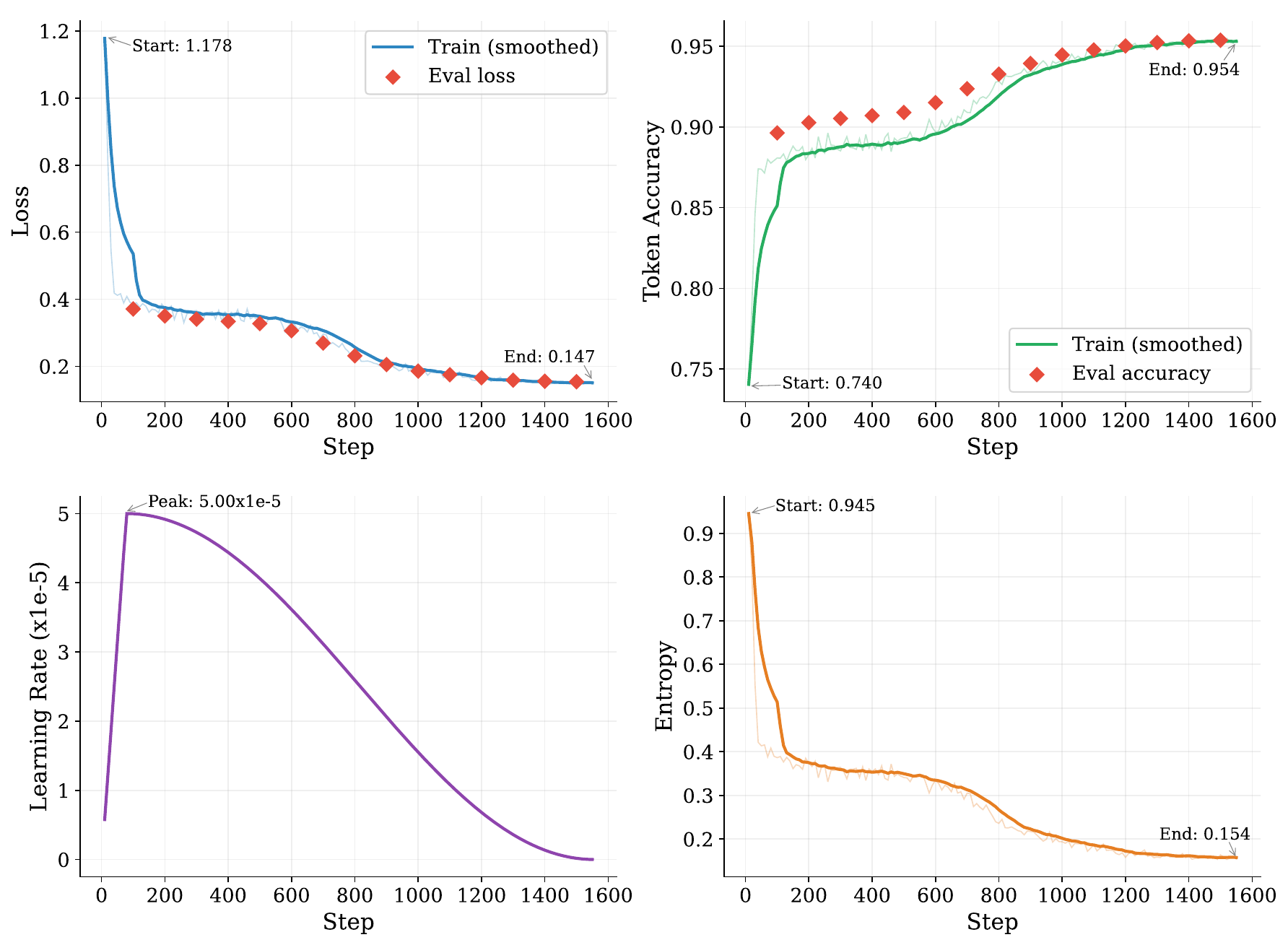} 
    \caption{Supervised Fine-Tuning (SFT) dynamics over 1,550 steps using the difficulty-sorted curriculum. \textbf{(A)} Training and evaluation loss, exhibiting an 87\% reduction from 1.178 to 0.147. \textbf{(B)} Token accuracy progression, rising from 0.740 to a final 0.954. The tight correlation between training and evaluation metrics throughout the run indicates minimal overfitting. \textbf{(C)} The learning rate schedule utilizing a cosine decay, peaking at $5.0 \times 10^{-5}$. \textbf{(D)} Output entropy decay, which drops sharply from 0.945 to 0.154, illustrating the model's increasing confidence in generating precise, structured malware analysis and vulnerability reports.}
    \label{fig:sft_dynamics}
\end{figure*}
%==============================================================

\textbf{Supervised Fine-Tuning:} \label{paragraph:sft_results} Following the initial domain adaptation, the SFT phase aligned the model to execute specific analytical workflows, with the progression detailed in~\cref{fig:sft_dynamics}. Utilizing a difficulty-sorted curriculum over 1,550 steps, the model demonstrated rapid instruction alignment, marked by an 87\% reduction in loss (from 1.178 down to 0.147). Crucially, the evaluation loss and accuracy tracked closely with the training metrics throughout the entire run, indicating that the model generalized well to the task structures without overfitting to the specific training templates. By the conclusion of the SFT phase, token accuracy reached 0.954. Additionally, the output entropy experienced a sharp decline from 0.945 to 0.154. This steep entropy reduction is a critical indicator of behavioral alignment, demonstrating that the model transitioned from generating diverse, exploratory tokens to producing highly confident, deterministic, and tightly structured threat intelligence reports.

%%%%%%%%%%%%%%%%%%%%%%%%%%%%%%%%%%%%%%%%%%%%%%%%%%%%%%%%%%%%%%%
% FIGURE: CORE TASKS SEMANTIC SIMILARITY
%%%%%%%%%%%%%%%%%%%%%%%%%%%%%%%%%%%%%%%%%%%%%%%%%%%%%%%%%%%%%%%
\begin{figure}[pos=htbp]
    \centering
    \includegraphics[width=\columnwidth]{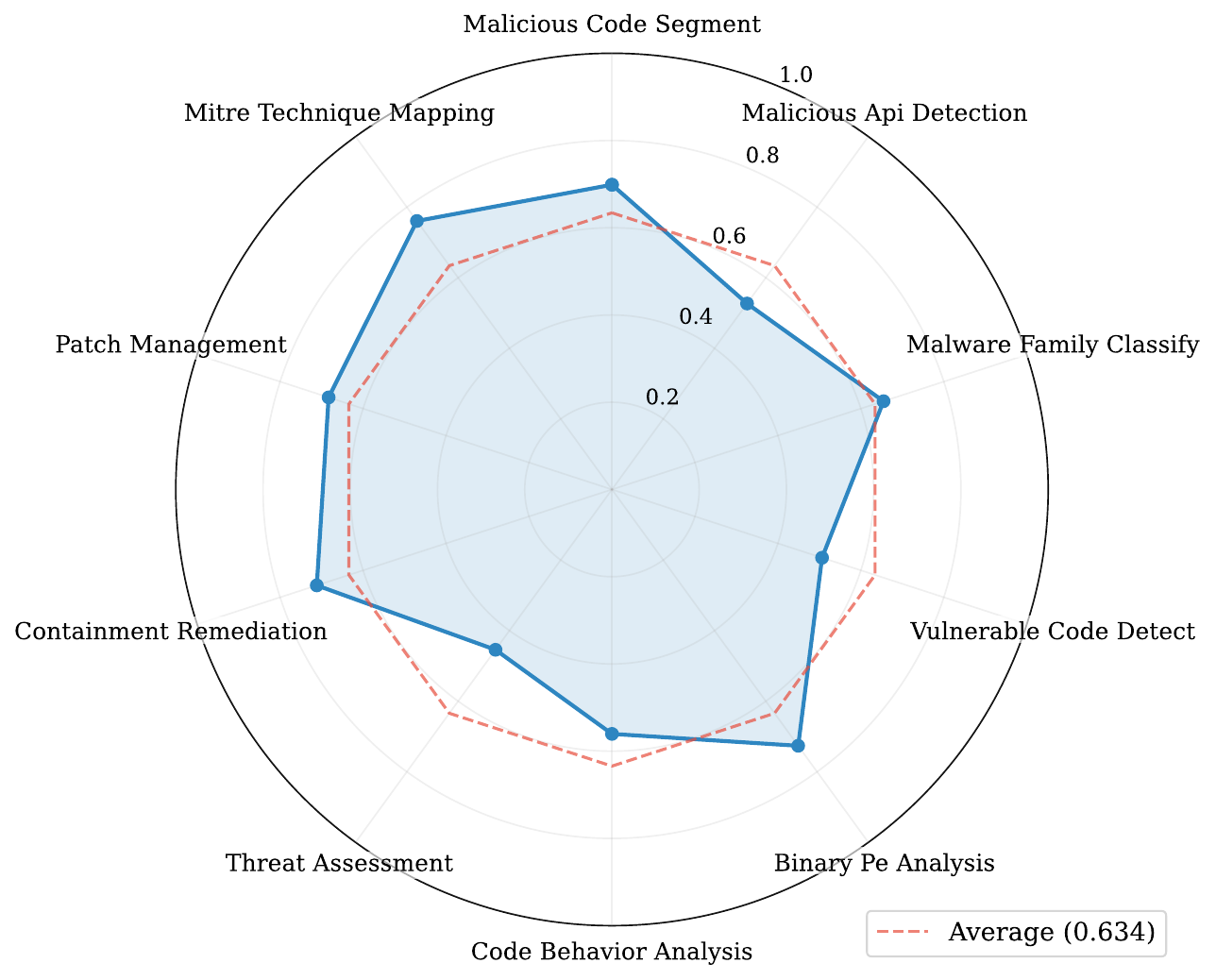} 
    \caption{Performance profile across 10 core malware-analysis tasks. Performance is measured via semantic alignment (SentenceTransformer cosine similarity) between model predictions and reference responses. The solid blue line represents the model's score on each specific axis. The dashed red baseline indicates the aggregate average similarity score ($0.634$) calculated across all 43 evaluated task types. Vertices extending beyond the baseline (e.g., MITRE ATT\&CK technique Mapping, Malware Family Classification) indicate relative strengths, largely driven by the model's pattern recognition capabilities. Points falling inside the baseline (e.g., Threat Assessment) highlight relative weaknesses in tasks requiring more nuanced reasoning.}
    \label{fig:task_radar}
\end{figure}
%==============================================================

%%%%%%%%%%%%%%%%%%%%%%%%%%%%%%%%%%%%%%%%%%%%%%%%%%%%%%%%%%%%%%%
% TABLE: EXACT-MATCH METRICS
%%%%%%%%%%%%%%%%%%%%%%%%%%%%%%%%%%%%%%%%%%%%%%%%%%%%%%%%%%%%%%%

\begin{table}[pos=htbp]
\centering
\scriptsize
\caption{Specialized Task Evaluation: Semantic Similarity vs. Exact-Match Metrics}
\label{tab:specialized_metrics}
\renewcommand{\arraystretch}{1.3}
\footnotesize
    \begin{tabular}{@{} >{\raggedright\arraybackslash}p{2cm} c >{\raggedright\arraybackslash}p{2cm} >{\raggedright\arraybackslash}p{2.4cm} @{}}
    \toprule
    \textbf{Task Type} & \textbf{Sim.} & \textbf{Specialized Metrics} & \textbf{Notes} \\ \midrule
    MITRE ATT\&CK technique Mapping & 0.761 & Precision: 0.526 \newline Recall: 0.357 \newline F1: 0.426 & Exact technique ID matching is harder than semantic alignment. \\ \addlinespace
    Malware Family Classify & 0.655 & Exact Accuracy: 0.0\% & Model provides relevant analysis even when label doesn't match exactly. \\ \addlinespace
    Vulnerable Code Detect & 0.506 & CWE F1: 0.0 & Needs more structured CWE training data. \\
    % \addlinespace
    % Malicious Code Segment & 0.699 & Token-Level F1: [PENDING] & Code extraction evaluated via token overlap to account for minor formatting variances. \\
    \bottomrule
    \end{tabular}
\end{table}
%==============================================================

\textbf{Task Specific Evaluation:} \cref{fig:task_radar} illustrates the mean semantic similarity across the 10 core malware-analysis tasks, providing a baseline measure of how well the fine-tuned model adopts the stylistic and structural conventions of expert malware analysis. While semantic similarity effectively captures the model's ability to reason and format responses appropriately, it is insufficient for tasks demanding rigid factual precision, such as classification tasks. A model can generate a highly convincing, structurally perfect analysis yielding high similarity while simultaneously failing to identify the exact classification or label that the task requires. To address this, we supplemented the semantic evaluation of~\cref{fig:task_radar} with other appropriate metrics for three core tasks requiring strict entity extraction, as detailed in~\cref{tab:specialized_metrics}.

These supplementary metrics reveal a contrast between structural alignment and factual precision. For instance, while MITRE ATT\&CK technique Mapping achieved a strong semantic similarity of 0.761, its F1 score was only 0.426, indicating difficulty in pinpointing exact T-codes despite understanding the broader attack mechanics. Similarly, Malware Family Classification reached a 0.655 similarity score despite a 0.0\% exact accuracy. In this specific task, the model consistently outputted the generic term ``classification'' rather than accurately identifying specific family names. Vulnerable Code Detection also struggled with exact retrieval, scoring a 0.0 F1 for specific CWE IDs. This divergence between metrics highlights a critical insight: while the model successfully learned the language and analytical flow of malware reports during training, achieving high reliability in extracting exact string segments, categorical IDs, or labels will require further targeted alignment.

\textbf{Complex Scenarios Evaluation:} To assess practical utility, the model was evaluated against five complex scenario prompts (full generated responses are provided in~\cref{sec:appendix:real-world_scenarios_responses}). Qualitative analysis of these responses revealed a strong adherence to CoT and CoVe reasoning patterns, which were deliberately augmented into the SFT dataset. In multiple scenarios, the model executed explicit, step-by-step reasoning preambles prior to outputting structured intelligence.

In the Family Classification scenario, this learned behavior was made explicitly visible by the model generating a ``\textless/think\textgreater'' token immediately before transitioning into its cleanly formatted markdown report. While this deliberate reasoning process enabled the model to correctly identify key indicators, the verbosity inherent to CoT and CoVe processing frequently caused the model to reach the token limit during inference, resulting in truncated final outputs. This indicates that while the SFT phase successfully aligned the model to utilize advanced reasoning and verification pathways, practical deployment will require either expanded inference context windows or targeted brevity constraints during training to accommodate these detailed analytical thought processes.

\section{Conclusion}\label{sec:conclusion}

We introduced LCCD and a tailored model training framework designed to advance the application of Large Language Models in automated malware analysis. While existing datasets offer fragmented views of malicious software, LCCD bridges this gap by unifying code-centric representations, static analysis artifacts, graph topologies, embeddings for RAG deployment, and curated instruction-response pairs into a single, cohesive resource. This combination equips researchers with the tools necessary to perform truly holistic analyses of malware behavior.

By incorporating these diverse representations, our training pipeline successfully facilitated the domain adaptation of a foundational LLM, enabling it to parse and reason over complex binary artifacts. While our fine-tuned model underperformed in tasks that required specific label identification or string extraction, it was able to produce accurate and well-structured reports that provide relevant insights for cybersecurity analysts. Ultimately, LCCD provides the open-source cybersecurity community with a robust foundation for benchmarking, pre-training, and deploying advanced AI-driven defense mechanisms.

\section*{Data Availability}

The Large-Scale Code-Centric Dataset used in this study is publicly available through the KAUST Repository at \url{http://hdl.handle.net/10754/709465}. The dataset can also be accessed using the DOI: \href{https://doi.org/10.25781/KAUST-Z05OK}{10.25781/KAUST-Z05OK}.

\bibliographystyle{unsrtnat}
% Loading bibliography database
\bibliography{references}
\newpage

\appendix
\crefalias{section}{appendix}

\section{Advanced Testing Scenario Outputs}\label{sec:appendix:real-world_scenarios_responses}
This appendix provides the raw, unedited outputs generated by the fine-tuned model during the testing phase. The responses highlight the model's adherence to Chain-of-Thought reasoning prior to generating structured reports. Note that several responses were truncated upon reaching the maximum token limit.

\subsection{Code Injection Analysis}
\noindent \textbf{Execution Time:} 514.2s | \textbf{Word Count:} 719 \\
\noindent \textbf{Response:}
\begin{quote}
\small
Okay, so I'm trying to understand this malicious DLL code that uses VirtualAlloc, CreateRemoteThread, and anti-debug techniques. I'm not super experienced with reverse engineering, but I'll do my best to break it down.

First, I know that VirtualAlloc is used to allocate memory in a process. In legitimate software, this is often used for things like allocating buffers or data structures. But in malware, it can be used for more nefarious purposes, like allocating space for stolen data or creating a buffer overflow to execute malicious code.

Next, CreateRemoteThread is a Windows API function that creates a thread in another process. This can be used by malware to inject code into a legitimate process, allowing it to execute malicious actions without being detected. I remember reading that this is a common technique used in banking trojans and other types of malware.

The anti-debug techniques in the code are designed to make it difficult for debuggers or reverse engineers to analyze the malware. Things like timing checks, API call patterns, and memory access patterns are used to detect when a debugger is attached. If the malware detects a debugger, it might terminate itself or take other actions to evade detection.

Looking at the code, I can see that it's using these APIs in a way that suggests it's trying to hide its presence. For example, the VirtualAlloc call is followed by a check to see if the memory allocation was successful. If it wasn't, the malware might terminate itself or take other actions to avoid detection.

The CreateRemoteThread call is also suspicious because it's creating a thread in another process without any obvious legitimate purpose. This could be used to inject malicious code into a legitimate process, allowing the malware to execute actions without being detected.

The anti-debug techniques are implemented using timing checks and API call patterns. For example, the malware might make a series of API calls in quick succession, which could be 
\end{quote}

\vspace{1em}

\subsection{Trojanized PE Detection}
\noindent \textbf{Execution Time:} 99.2s | \textbf{Word Count:} 82 \\
\noindent \textbf{Response:}
\begin{quote}
\small
\textbf{Step 1: File Triage}
File characteristics:

- \textbf{Size:} 3,392,815 bytes

- \textbf{Architecture:} x86\_64

- \textbf{Compiler:} MSVC 19.28

- \textbf{Entry point:} 0x820F

- \textbf{Functions detected:} 130

\textbf{Step 2: Code \& Behavior Analysis}
Examining decompiled code (273 functions) for:

- \textbf{Process manipulation:} No

- \textbf{Network API:} No

- \textbf{Registry API:} Yes

\textbf{Step 3: Indicator Identification}
Examining:

- \textbf{API usage patterns}

- \textbf{Behavioral indicators}

- \textbf{MITRE ATT\&CK mapping}

\textbf{Step 4: Assessment \& Classification}
Based on the above analysis:

\textbf{Classification:} AgentTesla (Downloader)

\textbf{Threat Level:} Low

\textbf{Malicious}

\end{quote}

\vspace{1em}

\subsection{Multi-Vuln SAST}
\noindent \textbf{Execution Time:} 546.8s | \textbf{Word Count:} 663 \\
\noindent \textbf{Response:}
\begin{quote}
\small
Alright, so I'm trying to analyze this C code for vulnerabilities. The code is part of a daemon, which means it's a background service running on the system. My goal is to find any buffer overflows, format string vulnerabilities, command injections, integer overflows, or double-free issues.

First, I'll look through the code to identify any obvious buffer overflow scenarios. Buffer overflows can occur when an application writes data beyond the bounds of a buffer, which can lead to code execution or data corruption. In C, this is common because of manual memory management.

Looking at the code, I see several instances of `memcpy`. For example, `memcpy(\&v1, \&v1, (int32\_t)\&v1);` This line seems suspicious because it's copying data from one location to another without checking the bounds. I should verify if the source and destination buffers are of appropriate sizes and if there's a risk of overlapping.

Next, I notice a `strcpy` function call: `strcpy((char *)str, (char *)str2);` This is another potential buffer overflow risk because it doesn't check the length of the source string. If `str2` is longer than `str`, this could lead to a buffer overflow.

I also see a `sprintf` function: `sprintf((char *)str, (char *)format, (int32\_t)v1);` This is a format string vulnerability because it uses user input (`v1`) in a formatted output function without proper validation. An attacker could potentially inject malicious code using this vulnerability.

Looking further, I find an instance of `malloc`: `malloc(0x40000000);` This is a memory allocation function, but I should check if the allocated memory is properly validated. If the allocation fails, it could lead to a null pointer dereference.

Another point of interest is the `free` function call: `free(NULL);` Freeing a null pointer is undefined behavior and can lead to crashes or security vulnerabilities. I should verify if this is intentional or a mistake.

I also notice some suspicious API calls, such as `CreateThread` and `O
\end{quote}

\vspace{1em}

\subsection{Family Classification}
\noindent \textbf{Execution Time:} 291.5s | \textbf{Word Count:} 356 \\
\noindent \textbf{Response:}
\begin{quote}
\small
Alright, let's tackle this malware classification task. The sample is a PE32+ executable with some suspicious imports and behaviors. First, I'll start by examining the file characteristics.

The file size is 3,111,896 bytes, which is quite large. Typically, larger files might contain more malicious code or be designed to evade detection through complexity. The architecture is x86, which is standard, but the presence of 128 functions might indicate a moderate level of complexity.

Looking at the imports, there are 20 suspicious ones. The top 3 by frequency are InternetConnectA, WSAStartup, and connect. These are related to network communication, which is a common feature in malware. The presence of CreateFileA and ReadFile suggests the malware might interact with files or network data.

Behavioral indicators include network activity, registry modifications, and process creation. The malware establishes outgoing connections, which could be for command and control (C2) communication. Registry modifications might involve creating or deleting keys, potentially for persistence or configuration storage.

The MITRE ATT\&CK technique mappings are T1071 (Application Layer Protocol) and T1547 (Privilege Escalation). T1071 is relevant due to the network communication indicators, while T1547 might not be directly applicable here unless there's evidence of privilege escalation.

Putting it all together, this malware sample exhibits characteristics typical of a keylogger with C2 beaconing capabilities. The large file size and moderate complexity suggest it's designed to evade detection through complexity and network communication.

\textless/think\textgreater

\textbf{Malware Classification Report}

\textbf{Sample:} PE32+ executable (keylogger imports, C2 beaconing, persistence via Run key)

\textbf{Classification:} Keylogger malware

\textbf{Threat Level:} Low

\textbf{Malware Family:} Keylogger

\textbf{Behavioral Indicators:}

* Network activity (establishing outgoing connections)

* Registry modifications (creating or deleting 
\end{quote}

\vspace{1em}

\subsection{Ransomware Containment}
\noindent \textbf{Execution Time:} 495.7s | \textbf{Word Count:} 719 \\
\noindent \textbf{Response:}
\begin{quote}
\small
Okay, so I'm trying to figure out how to contain this ransomware that's encrypting files on a shared drive. The user mentioned SMB lateral movement, so I guess the ransomware is spreading over the network. I'm not super experienced with incident response, but I'll try to break this down step by step.

First, I think I need to understand the current situation. What exactly are the indicators of this ransomware? Is it a known variant? I remember that ransomware often encrypts files with extensions like.encrypted,.locked, or specific cipher types. Maybe I should look for those patterns in the shared drive.

Next, I should assess the affected systems and network shares. Which machines are infected? Are there specific directories being targeted? It might help to map out the file system changes over time to see when the encryption started.

Then, network traffic analysis. I'm not sure how to do this without specific tools, but maybe I can look for unusual SMB activity---like multiple login attempts, directory traversals, or file transfers. If the ransomware is spreading, there might be connections to other machines.

Containment is the next step. I think isolating the infected machine is crucial to prevent further spread. Maybe disconnecting it from the network or blocking its IP address. Also, if the ransomware is using scheduled tasks or services, I should disable those to stop execution.

For eradication, I need to remove the ransomware. That means deleting the executable files, scripts, or services. I should also scan all affected machines with updated antivirus software to catch any remaining malware.

Recovering encrypted files is tricky. I don't know if the ransomware uses known ciphers, but maybe there are decryption tools available. I should check if the attackers provide a decryption key, but that's probably not reliable. Alternatively, maybe some files can be recovered from backups.

To prevent future incidents, I need to patch any vulnerabilities the ransomware 
\end{quote}

%*********************************************
\section{Real-Time LCC-LLM Chatbot Prototype}
\label{app:chatbot_prototype}
A real-time chatbot prototype was developed to demonstrate the practical deployment of LCC-LLM as an analyst-facing malware-analysis expert system. The chatbot allows analysts to interactively query malware samples and request explanations related to malware attribution, malicious code segments, suspicious APIs, IoCs, MITRE ATT\&CK mappings, vulnerability analysis, and containment guidance. The chatbot is connected to the proposed LCC-LLM pipeline, where retrieved cybersecurity knowledge, static-analysis evidence, verification, and quality-gated reasoning are combined to generate structured responses.
\begin{figure*}[t]
    \centering
    \includegraphics[width=\textwidth]{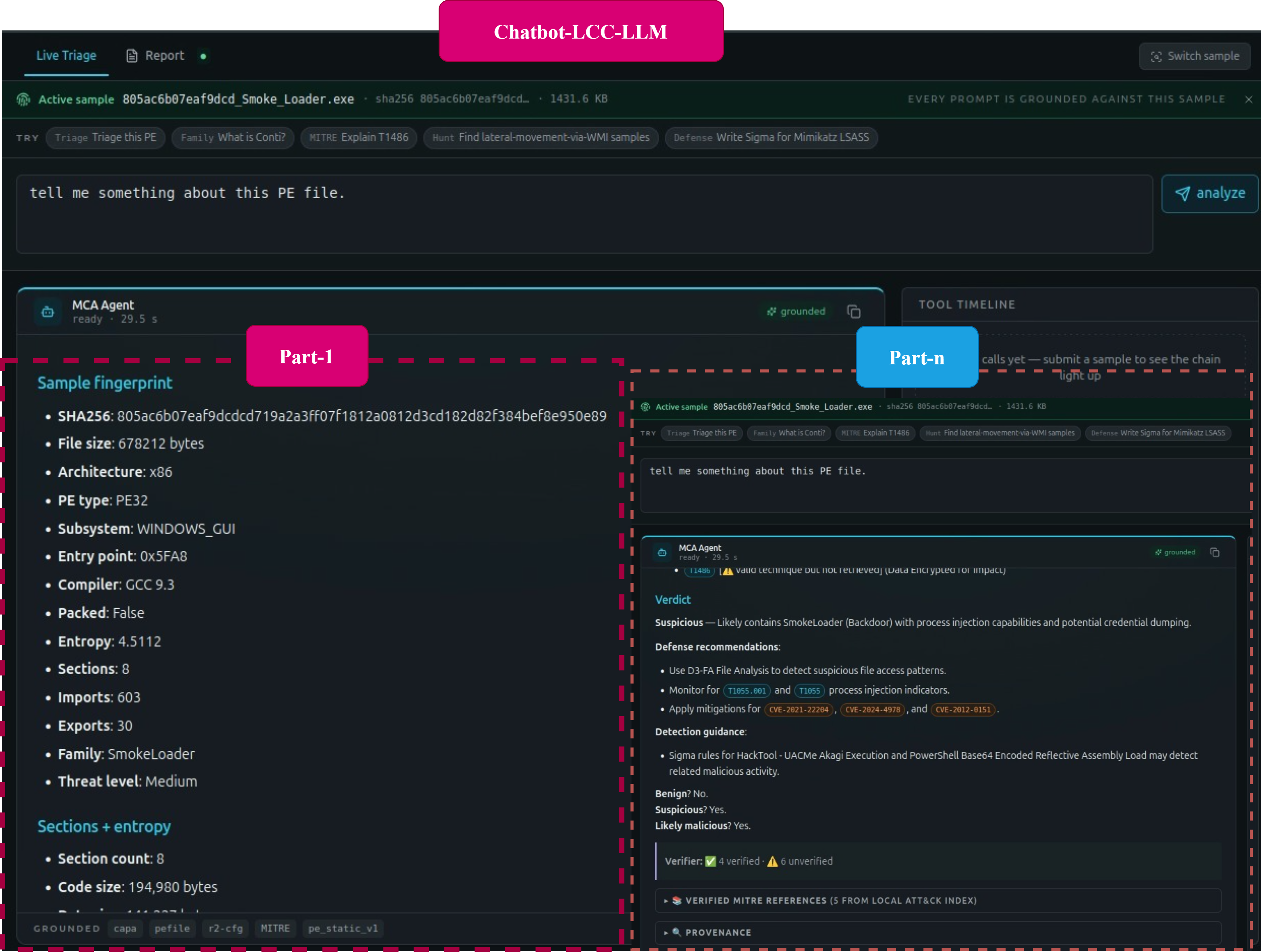}
    \caption{Real-time LCC-LLM chatbot prototype for interactive malware triage and analyst-oriented malware attribution.}
    \label{fig:chatbot_interface_1}
\end{figure*}

\begin{figure*}[t]
    \centering
    \includegraphics[width=\textwidth]{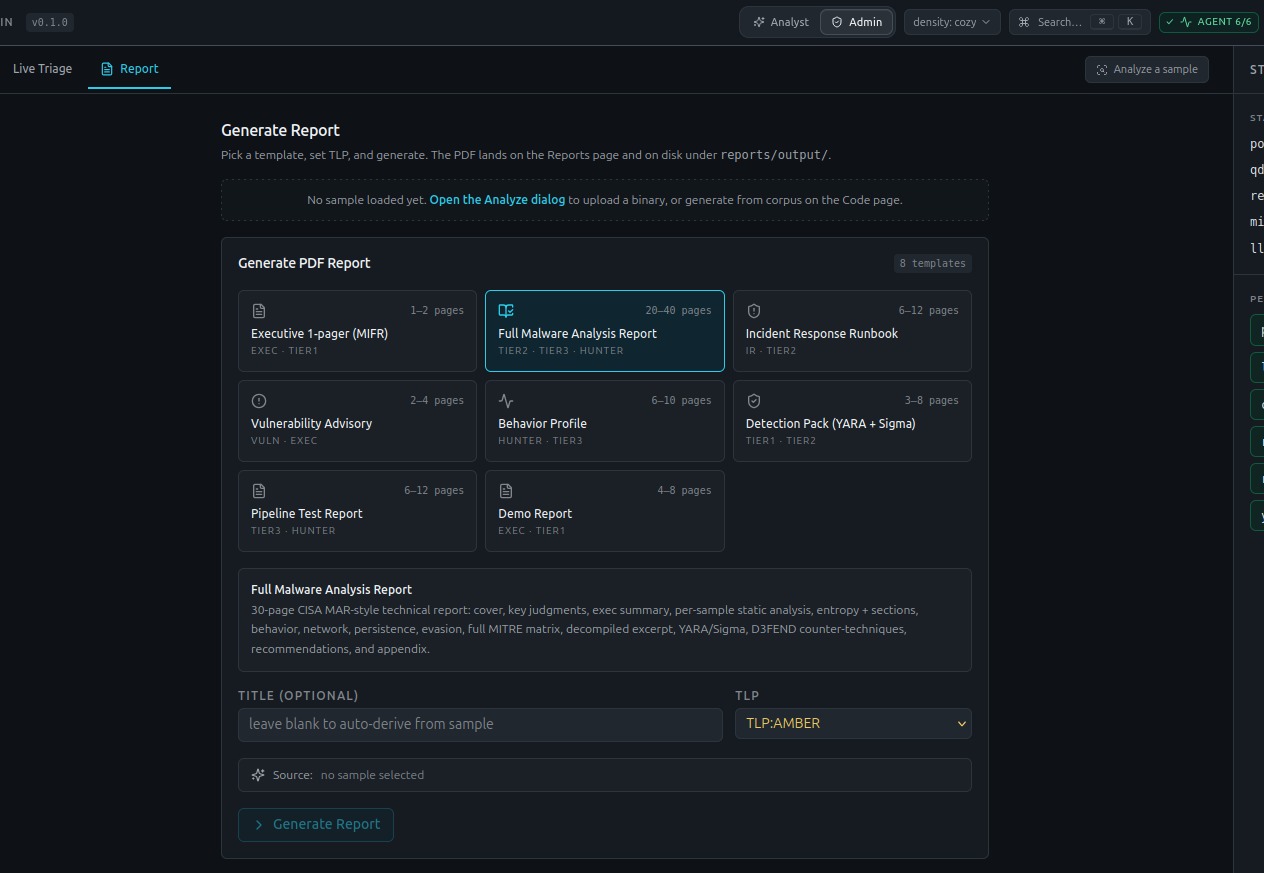}
    \caption{Real-time LCC-LLM chatbot prototype for interactive malware triage and analyst-oriented malware attribution.}
    \label{fig:chatbot_interface_2}
\end{figure*}
% Uncomment and use as the case may be
%\begin{theorem} 
%\end{theorem}

% Uncomment and use as the case may be
%\begin{lemma} 
%\end{lemma}

%% The Appendices part is started with the command \appendix;
%% appendix sections are then done as normal sections
%% \appendix
% \section{}\label{sec:}

% To print the credit authorship contribution details
%\printcredits

%% Loading bibliography style file
%\bibliographystyle{model1-num-names}

% Biography
%\bio{}
% Here goes the biography details.
%\endbio

%\bio{pic1}
% Here goes the biography details.
%\endbio

\end{document}